%% file: MousaviCSG_twc.tex
\documentclass[12pt, draftclsnofoot, onecolumn]{IEEEtran}

\usepackage{etex}
\ifCLASSINFOpdf
   \usepackage[pdftex]{graphicx}
\else
   \usepackage[dvips]{graphicx}
\fi
\usepackage[noadjust]{cite}
\usepackage{filecontents}
\usepackage[cmex10]{amsmath}
\usepackage{array}
\usepackage{svg}
\usepackage{url}
\usepackage[normalem]{ulem}
\usepackage{psfrag}
\usepackage{dsfont}
\usepackage{amssymb}
\usepackage{textcomp}
\usepackage{algpseudocode}
\usepackage{pifont}
\usepackage{mathtools}
\usepackage[cmex10]{amsmath}
\usepackage{amsfonts}
\usepackage{xcolor}
\usepackage{epstopdf}
\usepackage{amsmath,epsfig,amssymb,mathrsfs,psfrag,epsf,enumerate,bm,color}
\usepackage{booktabs}
\usepackage{color}
\usepackage{tikz}
\usepackage{graphicx}
\usepackage{lipsum}
\usepackage{amsthm}
\usepackage{soul}
\usetikzlibrary{shapes,arrows,automata,positioning}

\usepackage[usenames,dvipsnames]{pstricks}
\usepackage{epsfig} 
\usepackage{pst-grad} 
\usepackage{pst-plot} 
\usepackage{chngcntr}
\usepackage{wrapfig}

\usepackage{float}
\usepackage{multicol}
\usepackage[font=footnotesize,labelfont=bf]{caption}
 
\usepackage{epstopdf} 
\usepackage{setspace}

\usepackage{makecell} 
\usepackage{hyperref}
\usepackage{algorithm}
\usepackage{cleveref}

\usepackage{array}
\usepackage{setspace}

\newtheorem{theorem}{Theorem}
\newtheorem{lemma}{Lemma}
\newtheorem{definition}{Definition}
\newtheorem{observation}{Observation}
\newtheorem{corollary}{Corollary}
\newtheorem{remark}{Remark}

\newcommand{\argmin}{\operatornamewithlimits{argmin}}

\newcommand*\squared[1]{\tikz[baseline=(char.base)]{
            \node[shape=rectangle,draw,outer sep=2pt] (char) {#1};}}

\makeindex

\begin{document}
 
\title{ Cost Sharing Games for Energy-Efficient Multi-Hop Broadcast in  Wireless Networks }
 
\author{
\IEEEauthorblockN{
	Mahdi Mousavi,~\IEEEmembership{Student Member,~IEEE,}
	Hussein Al-Shatri,~\IEEEmembership{Member,~IEEE,}
	and 
	Anja Klein,~\IEEEmembership{Member,~IEEE,}   
     }
 \thanks{The authors are with Communications Engineering Lab, Technische Universit\"at Darmstadt, Germany.
E-mail: \{m.mousavi,  h.shatri, a.klein\}@nt.tu-darmstadt.de }
}

\maketitle
 
\begin{abstract} 
We study multi-hop broadcast in  wireless networks with one source node and multiple receiving nodes. 
The message flow from the source to the receivers can be modeled as a tree-graph, called broadcast-tree. 
The problem of finding the minimum-power broadcast-tree (MPBT) is NP-complete. 
Unlike most of the existing centralized approaches, we propose a decentralized algorithm, based on a non-cooperative cost-sharing game. 
In this game, every receiving node, as a player, chooses another node of the network as its respective transmitting node for receiving the message. 
Consequently, a cost is assigned to the receiving node based on the power imposed on its chosen transmitting node. 
In our model, the total required power at a transmitting node consists of (i) the transmit power and (ii) the circuitry power
needed for communication hardware modules. 
We develop our algorithm using the marginal contribution (MC) cost-sharing scheme and show that 
the optimum broadcast-tree is always a Nash equilibrium (NE) of the game.
Simulation results demonstrate that our proposed algorithm outperforms conventional algorithms for the MPBT problem. 
Besides, we show that the circuitry power, which is usually ignored by existing algorithms, significantly impacts the energy-efficiency of the network.
\end{abstract}

\IEEEpeerreviewmaketitle

\section{ Introduction} 
\label{sec:intro}

This paper focuses  on the energy-efficiency of  multi-hop broadcast in wireless  networks  where a source  has a common message for a number of other nodes.
The source's message is disseminated through the network with the help of some intermediate nodes which re-transmit the message.
This problem is known as the minimum-power broadcast-tree (MPBT) problem since the connections between the source and the receiving nodes form  a tree-graph rooted at the source, called the \textit{broadcast-tree} (BT) \cite{BIP}.   
MPBT construction has been studied by researchers extensively during the past two decades \cite{BIP, EWMA,  Caragiannis_TON13,  AntColony_Spr09, PSO_Els13,  GA_Els11, DynaBIP, Cartigny_infocom03, Rahnavard08,  Komlai_TMC_08, Chen_TMC13, Chekuri:JSAC07}.
NP-completeness of the MPBT can be shown   by reducing the Steiner tree problem to it \cite{Liang_MobHoc02_NPCom}. 
This means that  a polynomial-time algorithm to find the optimum BT unlikely  exists. 
Although many algorithms have been proposed for the MPBT problem, most of them are \textit{centralized} heuristics \cite{BIP, EWMA,  Caragiannis_TON13,  AntColony_Spr09, PSO_Els13,  GA_Els11 }. 
Since multi-hop device-to-device communication is seen as a promising  technique for improving the capacity of 5G cellular networks \cite{Tehrani_SM14} and due to the variety of its applications, e.g., video streaming \cite{Baccouch_wcnc17}, vehicular communications \cite{Lai_ComLetter18},   etc., it is vital to revisit the MPBT problem to find a \textit{decentralized} yet efficient algorithm for it.

Seeing the MPBT problem from a decentralized optimization point of view, every individual node has to play its own role in forming the BT by establishing a communication link to another node.
This can be suitably modeled by game theory, in which the players of the game, here the nodes, are typically modeled as selfish agents seeking to minimize (maximize) their own cost (revenue).
In our work, the action of a node is to choose another node in the network as its respective transmitting node to receive the source's  message from.
As a consequence of its decision, a cost is assigned to the node based on the power it imposes on its chosen  node.

In a game in which a group of players benefits from one resource, the cost of using the resource (the transmit power required at a transmitting node) can be shared among the ones who use it (nodes of a multicast receiving group).
This class of game is called   cost sharing game (CSG) \cite{MAS08}.
Using a CSG, the nodes are motivated to choose a common transmitting node  which can lead to network power reduction by reducing the  number of transmissions.
In our proposed game, the goal of every node is to minimize its own cost and hence, the game is called a non-cooperative cost sharing game.
The Nash equilibrium (NE)  is considered as the solution concept of the game at which none of the nodes changes its chosen transmitting node.
To reach an NE, we employ the so-called \textit{best response dynamics} by  which  the nodes iteratively update their actions and at every iteration a node "best-responds" to the actions taken by the other nodes.
We show that our proposed game is an \textit{ exact potential game} \cite{MAS08} for which an NE is always guaranteed to exist.
 Moreover, we show that with our proposed game, the optimum state of the BT is always an  NE of the game.

The rest of the paper is organized as follows: 
Section \ref{sec:relatedandcontrib} reviews the related works and explains the contribution of our work.
The network model and problem statement are presented in  Section \ref{sec:netmodel}. 
Our game-theoretic algorithm is explained in Section \ref{sec:game}. 
In Section \ref{sec:MILP}, a centralized approach is provided for the MPBT problem as a benchmark for our proposed algorithm. 
The simulation results are presented in Section \ref{sec:simul} and Section \ref{sec:conc} concludes the paper.

\section{Related Work and Contribution } 
\label{sec:relatedandcontrib}
\subsection{Related Works} \label{subs:relatedw}
The algorithms  proposed for the MPBT  are usually not able to find the optimum BT, especially when the number of nodes in the network is large, but they can find a low power BT in  polynomial-time.
A well-known heuristic   called the broadcast incremental power (BIP) algorithm is proposed in \cite{BIP}.
The BIP algorithm is a  centralized greedy heuristic. 
To build a BT, it starts from the source and  iteratively connects the nodes  to the source or to the other nodes already connected to the BT.
Considering the transmit power of the nodes which are already connected to the BT, in each iteration, the node which requires the minimum  incremental transmit power is chosen as the new node to connect to the BT \cite{BIP}.
Since the BIP algorithm fails in exploiting the benefit of multicast transmission in the wireless medium, the authors of \cite{BIP} further  propose  a procedure called sweeping in order to improve  their  algorithm.
We refer to  the BIP algorithm along with the sweeping procedure as the BIPSW algorithm.
When the BT is initialized by the BIP algorithm, the BIPSW prunes the links to the nodes which can be covered by other transmitting nodes and  prevents unnecessary transmissions.
Other heuristics based on minimum spanning tree \cite{EWMA, Caragiannis_TON13}, ant colony optimization \cite{AntColony_Spr09}, particle swarm optimization \cite{PSO_Els13}, and genetic algorithm \cite{GA_Els11}   have also been proposed during the past years for the MPBT problem which all are centralized and may perform better than the BIP algorithm at the expense of a higher complexity.
 
The main drawback of the centralized algorithms is their dependency on an access point.
This makes the network vulnerable if the nodes lose their connections to it.
Hence, decentralized algorithms \cite{Rahnavard08, DynaBIP, Cartigny_infocom03}, by which the nodes construct the BT just based on their local information, are a better choice for real-world implementations. 
Since in a decentralized algorithm  the nodes update their action  independently, to find a valid tree-graph as BT,  the algorithm may require to be initialized to restrict the decisions of the nodes.
The authors of \cite{Rahnavard08} suggest an algorithm called the broadcast decremental power (BDP) which first  initializes the BT by a centralized algorithm (Bellman-Ford), and then, every node changes its respective transmitting node if the change leads to a lower transmit power. 
A decentralized algorithm is also suggested in  \cite{Cartigny_infocom03},  but it requires the geographical position of all the nodes of the network at every single node.
Decentralized approaches for the MPBT problem have received less attention, and in general, lack a good performance compared to the centralized ones.

Game theory, as a powerful mathematical tool, has  been widely  used for  designing games for distributed optimization \cite{marden_welfare13, Chekuri:JSAC07, Chen_TMC13, Mousavi_ISWCS15} or resource sharing in  competitive situations \cite{GameSurvey16,  Haji_JSAC17}. 
For instance, the authors of \cite{Komlai_TMC_08} exploit a potential game to control the topology and maintain the connectivity of a multi-hop wireless network.
Their proposed approach does not consider multicast transmission and requires the information from several hops to be collected at every single node. 
Different cost sharing schemes, each with different properties in terms of implementation difficulties or convergence to an NE, can be employed in a CSG to share the cost of a multicast transmission among the nodes.
The authors of \cite{SA11} studied some of the schemes that can be used for coalition formation for a single-hop multicast transmission.

In \cite{Mousavi_ISWCS15} and \cite{Alex15}, we showed that a  CSG is a suitable decentralized approach to be used for modeling the MPBT problem. 
An important class of  sharing schemes for  CSGs is the class of budget-balanced schemes \cite{Gopalakrishnan:potgame:14, Chen:2010:SIAM:GoodEq, Dobzinski2008}.
A cost sharing scheme is budget-balanced if the sum of the cost allocated to each of the receiving nodes  of a multicast transmission (entities involved in a coalition) is equal to the transmit power of the transmitting node (the price of the resource used by the entities).
One of the widely-adopted budget-balanced schemes is the Equal-share (ES) scheme in which the cost is simply shared equally among the nodes.
Due to difficulties of designing a decentralized approach to the MPBT problem, the simplified versions of this problem have also been studied in the literature which we call the minimum-transmission BT (MTBT) and the minimum fixed-power BT  (MFPBT) problems.
In the MFPBT problem, the nodes have fixed but not necessarily equal transmit powers while in the MTBT problem, the transmit powers  of the nodes are not only assumed to be fixed  but also  \textit{equal} for all the nodes.
In fact, the MTBT problem  is a special case of the MFPBT problem and both of them are   simplified versions of the MPBT problem.
The ES cost sharing scheme has been employed in \cite{Chen_TMC13} and \cite{ Chekuri:JSAC07 } for the MTBT and MFPBT problems, respectively.
The algorithm in \cite{Chen_TMC13} is called GBBTC and the authors, by assuming a fixed transmit power at the nodes, minimize the number of transmissions in the network as a way to minimize the network power.

The GBBTC algorithm studied in \cite{Chen_TMC13} has two main drawbacks.  
Firstly, it does not perform power control at the transmitting nodes. 
Secondly, the ES cost sharing scheme employed in   \cite{Chen_TMC13} (also in \cite{Chekuri:JSAC07}), is not applicable to the MPBT problem since the convergence of the state of the BT to an NE cannot be guaranteed.
In fact, as we will show, to ensure the convergence to an NE when using the ES cost sharing scheme, the power control feature at the nodes cannot be exploited and a fixed transmit power must be used instead.
Besides addressing these drawbacks in our proposed algorithm, we use a power model for the nodes which is more realistic than the models  commonly-used in the literature  \cite{BIP, EWMA, DynaBIP, Cartigny_infocom03, Caragiannis_TON13,  AntColony_Spr09, PSO_Els13, Rahnavard08, GA_Els11, Komlai_TMC_08, Chen_TMC13, Chekuri:JSAC07} and show that with the proposed model the energy-efficiency of the network can be significantly improved.

\subsection{Our Contribution} \label{subs:contrib} 
Despite its wide adoption, to the best of our knowledge, game theory has not been  used  for the MPBT problem in which the nodes can perform transmit power control.
We propose a  CSG for the MPBT problem based on the MC cost sharing scheme, in short MC, and refer to it  as CSG-MC.
We further study two of the well-known budget-balanced schemes, the ES   and the Shapley value (SV), and compare their properties with the MC for the MPBT and the MTBT problems.
We will show that the scheme based on which the cost is shared among the receiving nodes of a multicast transmission significantly impacts the performance of the game and its convergence to an NE.
Although the MC   is not budget-balanced, we will show that it is the cost sharing scheme for which  the local objective at the nodes (cost minimization) is exactly aligned with the global objective (network power minimization).
This vital property does not hold, in general, for  budget-balanced cost sharing schemes for the MPBT problem.
We also show that,  with the MC, the optimum state of the BT is always an NE.

In the present work, we consider a more general power model than commonly-studied models in multi-hop networks \cite{Komlai_TMC_08, Chen_TMC13, Alex15}. 
Our proposed cost model consists of both transmit power for the radio link and  circuitry power of a transmitting node as the total power required at a transmitting node. 
While most of the existing works ignore the circuitry power of wireless devices and just focus on the power required for the radio link,  the circuitry power imposed on a transmitter  has a significant impact on the energy consumption in a wireless network \cite{Auer_WC11}.
In practice, not only the circuitry power  is  not negligible compared to the transmit power required for the radio links  but also it can dominate  when the distance between the transmitter and receiver is short \cite{Cui_TWC05, wang_SECON06_sht}.  
For instance, if the network is dense, having a single-hop broadcast would be more energy-efficient than having multiple short hops.
Note that the impact of circuitry powers cannot be seen as a fixed value on top of  the result obtained by an algorithm that ignores the circuitry power. 
 In fact, as we will show, taking the circuitry power into account may significantly change the structure of the BT and having an algorithm that captures both the device's circuitry power and radio link power jointly in BT construction is of high importance.  
Table \ref{tab:related} summarizes the main differences between our work and the benchmark algorithms discussed earlier.

\begin{table*} [!t] 
	\centering
  	\caption{ Comparison between different  algorithms proposed   for the MPBT  problem.  } 
  \small
\begin{center}
\resizebox{\textwidth}{!}{ 
 \begin{tabular}{|c||c|c|c|c|c|} 
 \Xhline{2.2 \arrayrulewidth}
   & BIPSW \cite{BIP} & BDP \cite{Rahnavard08}  & MFPBT \cite{Chekuri:JSAC07} & GBBTC \cite{Chen_TMC13} & CSG-MC \\
  & & & & & (This work)  \\
 \hline
 \Xhline{2.2 \arrayrulewidth}  
  \addlinespace[.5mm]
 \Xhline{2.2 \arrayrulewidth}  
Decentralized &  \ding{53} & \ding{51} & \ding{51} & \ding{51} & \ding{51} \\ \hline
Transmit power control &  \ding{51} & \ding{51} & \ding{53} & \ding{53} & \ding{51}\\   \hline
Circuitry power consideration &  \ding{53} & \ding{53} & \ding{53} &  \ding{53} & \ding{51}\\ \hline
Different  max. transmit power &  \ding{51} & \ding{51} & \ding{51}  & \ding{53} & \ding{51} \\   
\hline 
\Xhline{2.2 \arrayrulewidth}
 \end{tabular}
}
\end{center}
\label{tab:related}
\end{table*}

Finally, as the main  benchmark for  our algorithm, we formulate  the MPBT problem  as a mixed integer linear optimization problem (MILP).
Although   MILP formulations of the MPBT problem have been proposed in the literature, they usually do not take the circuitry power of the nodes into account.
Moreover, they mostly rely on finding the optimum value of the MPBT and do not explicitly suggest how to construct the BT \cite{Das_infocom03, Montemanni_WCNC05, abc_2004}.
Beside the MILP formulation of the MPBT problem, we provide a pseudo-code and explain how the optimum BT can be constructed using the output of the proposed MILP. 
 Notice that,  since  the MFPBT and MTBT problems are special cases of the MPBT problem, our proposed game  and the MILP formulation can also be applied to those problems.

 In summary, the main contributions of this paper are as follows:
 \begin{itemize}
 \item 
 We propose a decentralized game-theoretic algorithm for the MPBT problems  using the MC cost sharing scheme that can also be applied to the MFPBT and MTBT problems. 
 With our algorithm, while the transmitting nodes can perform power control, the convergence of the game to an NE is guaranteed.

\item
We extensively discuss the properties of our MC-based algorithm and  two other budget-balanced cost sharing schemes.
We show that with MC, unlike budget-balanced CSCs, the optimum BT is always an NE.
Moreover, by using a budget-balanced scheme, the network power minimization may be in contrast to the node's cost minimization.
We further show that  the ES does not guarantee the existence of an NE for the MPBT problem.

 \item 
 The proposed algorithm captures both transmit power and device's circuitry power jointly in BT construction. 
 To the best of our knowledge, considering the circuitry power in BT construction has not been studied before.
 We show that the device's circuitry power has a significant impact on the energy efficiency of the network.  
 
 \item 
 An MILP formulation is proposed for the MPBT problem considering the device's circuitry power as well as an algorithm that constructs the optimum BT.
 \end{itemize}

\section{Network Model and Problem Statement} \label{sec:netmodel}

A network composed of $N+1$ wireless nodes with random locations in a two-dimensional plane is considered; a source $\mathrm{S}$ and a set  $\mathcal{W} $ of $N$ receiving nodes.
The nodes in $\mathcal{W}$ are  interested in receiving the source's message.
We denote the set of all nodes of the network as $\mathcal{Q} = \mathcal{W} \cup \{\mathrm{S}\}$.
Every node is equipped with an omnidirectional  antenna and has a  \textit{transmit} power constraint $p^\mathrm{max}_j , j\in \mathcal{Q}$, and hence, its coverage area is limited.

\begin{figure}[t]
   \centering
   \includegraphics[width=.4 \columnwidth]{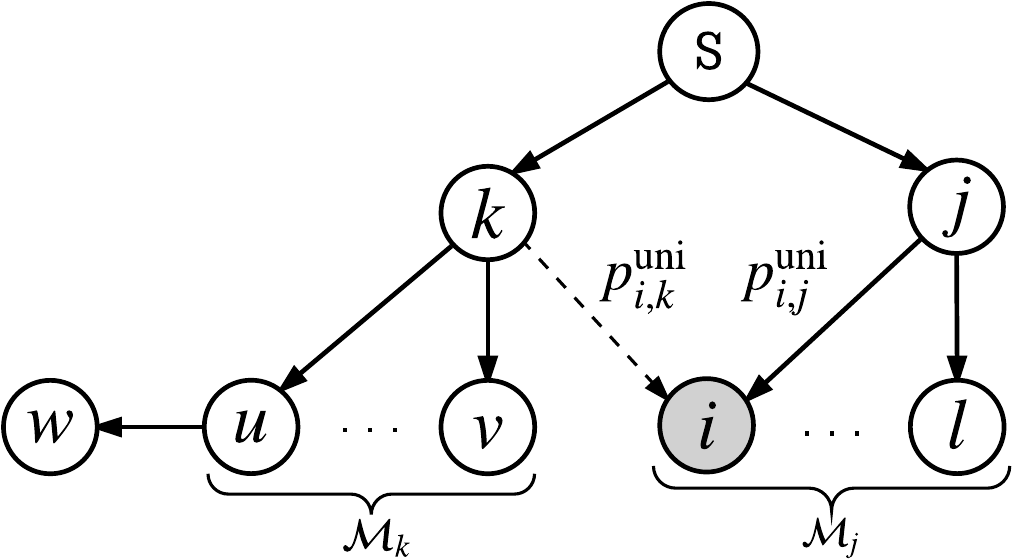}
   \caption{\footnotesize A sample network and BT.  The solid  arrows show the transmission from PNs to CNs and the dashed arrow  represents a possible link that can be established for CN  $i$ if it chooses PN $k$.   }
   \label{fig:net}
\end{figure}

In a transmission from  a transmitting node $j \in \mathcal{Q}$ to a receiving node $i \in \mathcal{W}$, nodes $j$ and $i$ are called  the \textit{parent node} (PN) and the \textit{child node} (CN), respectively.
The transmitting nodes transmit either by multicast or unicast.
It should be remarked that, although the antenna broadcasts the message omnidirectionally,  we  refer to the transmission as   unicast or multicast, when a PN has one or more than one intended receivers as its CNs, respectively. 
A CN receives the message from its own PN and ignores the messages transmitted by the other nodes.
The set of CNs of PN $j$ is denoted by $\mathcal{M}_j$, see Fig. \ref{fig:net}.
It is assumed that every CN is served by just one PN and if a node $j \in \mathcal{W}$ does not forward the message to any other node, then $\mathcal{M}_j =\varnothing$.
 
The hardware of a wireless device is composed of several modules such as base-band signal processing  unit, digital-to-analog  converter, power amplifier, etc., where every component has its own power requirement  for proper operation \cite{Cui_TWC05}.
The total power required at a transmitting node consists of two main parts.
The first part is the power required for the modules that mainly prepare the signal for transmission.
We refer to this first part as the \textit{circuitry} power of the node and denote it by $p_j^\mathrm{c}, \forall j \in \mathcal{Q}$.
The second part is the power that has to be spent by a transmitter to amplify the signal,  referred  to as the \textit{transmit} power of a node.
As mentioned before, the circuitry power of a wireless device is not negligible compared to the transmit power and may even dominate  it if the distance between the transmitter and the receiver is short \cite{Cui_TWC05}.
Hence, we assume that every node $j \in \mathcal{Q}$ has a total power budget of $p_j^\mathrm{c} + p_j^\mathrm{max}$.
While the circuitry power of a transmitter can be assumed as a fixed value, the transmit power needed at a transmitting node $j$ for amplifying the signal  depends on the channel quality between the transmitter  and its receivers in $\mathcal{M}_j$ and thus, it is denoted by $p_j^\mathrm{Tx}(\mathcal{M}_j)$ with $p_j^\mathrm{Tx}(\mathcal{M}_j) \leq p_j^\mathrm{max}$.
The \textit{total} power required at \textit{node} $j$ for transmission to its CNs in $\mathcal{M}_j$ is  
\begin{equation}
\label{eq:node_pow_def}
P_j (\mathcal{M}_j)= p_j^\mathrm{c} +  p_j^\mathrm{Tx}(\mathcal{M}_j). 
\end{equation}
We refer to the PN of CN $i$ as $a_i$ such that $a_i = j, j \in \mathcal{Q}\backslash \{i\}$ and $\mathbf{a} = (a_1, \dots, a_N)$ represents a vector whose elements are the PNs of each of the nodes in $\mathcal{W}$.
For the sake of notational convenience, we use $P_j (\mathbf{a})$ instead of  $ P_j (\mathcal{M}_j)$, when required.
Note that we omit the circuitry power required for message reception as it does not affect the energy-efficiency of the network.
In other words, circuitry power required for receiving data, usually a fixed value, is needed at every node that aims to receive  the message and this energy does not depend on the BT.

 The power $p_j^\mathrm{out}$  of the signal emitted from the antenna of a transmitter  $j$ depends on the efficiency of its  power amplifier, denoted by $\eta_j$ with $ 0<\eta_j<1$,   as is given by \cite{wang_SECON06_sht}
\begin{equation} \label{eq:Pamplif}
p_j^\mathrm{out} = \eta_j p_j^\mathrm{Tx}.
\end{equation}
For the message reception, a threshold model is considered  in the network, that is, a minimum signal-to-noise ratio (SNR), denoted by $\gamma^\mathrm{th}$, is required at a CN  for successful reception of the message transmitted from its PN.
In other words, the bit-error rate is assumed to be negligible considering $\gamma^\mathrm{th}$. 
We assume that the statistical properties of the channel remain invariant during the data transmission.
Let $g_{i,j}$ be the channel gain between the PN $j$ and the CN $i$. 
By treating interference as noise and denoting their joint power at the receiver  $i$ as $\sigma^2$, the SNR of the signal received by  CN $i$   is given by  
\begin{equation}
\label{eq:snrdef}
\gamma_{i,j} = \frac{ p_j^\mathrm{out} g_{i,j}  }{ \sigma^2 }.  
\end{equation}
Based on  the minimum required SNR $\gamma^\mathrm{th}$ at CN $i$, and using \eqref{eq:Pamplif} and \eqref{eq:snrdef}, the transmit power required at a transmitting node $j$ for transmission to a receiving node $i$  is given by
\begin{equation}
\label{eq:puni}
p_{i,j}^\mathrm{req}  =  \frac{\gamma^\mathrm{th}  \sigma^2 }{ \eta_j g_{i,j} }.
\end{equation} 
 Notice that  $p_{i,j}^\mathrm{req}$   takes the efficiency of the power amplifier of the transmitting node into account.

Our algorithm can   employ any decentralized channel access scheme suitable for multi-hop communications.
For instance,  a  time-slotted channel can be used that  consists of two sections, the first section as random access and the second section as scheduled access.
The first section is contention-based and used for signaling message  exchanges  while the transmissions by the PNs are carried out during the scheduled access section.
Such a model has been studied  and adopted by standards  during the past years \cite{Rajendran:collFree:06,  Tavana:ICC:15, 3GPP:rpt:rach:10}.

\begin{figure}[t]
   \centering
   \includegraphics[width=.65 \columnwidth]{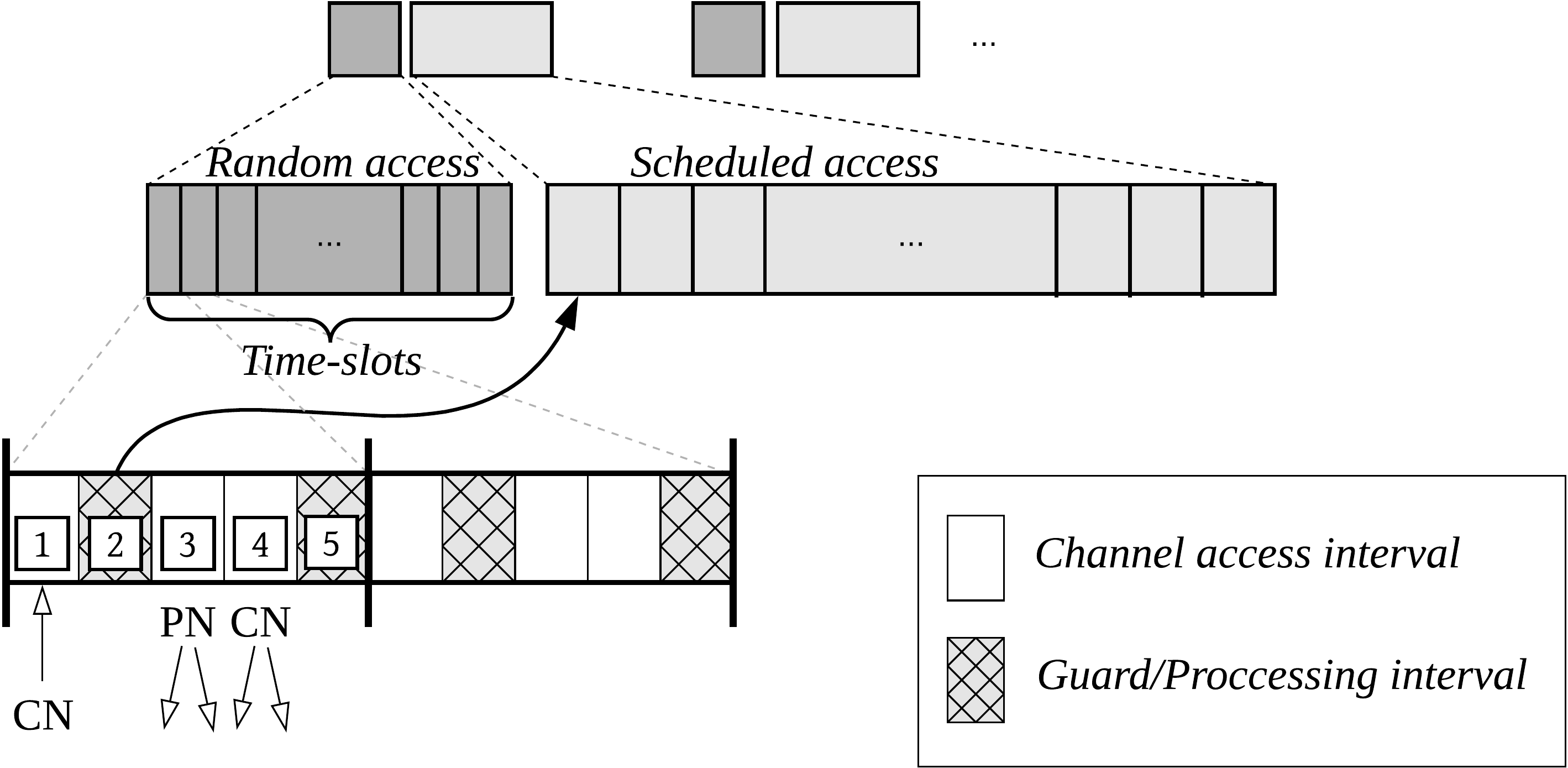}
   \caption{\footnotesize Channel access scheme.   }
   \label{fig:chaccess}
\end{figure}

Fig. \ref{fig:chaccess} shows a model for channel access.
In this model, the random access channel (RACH) phase, i.e., the first  phase, is divided into several time-slots, each divided into  five main intervals.
\\
Let us assume that a node $i$ decides to access the channel in the random access phase in time-slot $s$.
Node $i$ sends its request  to node $j$ in the first interval of time-slot $s$, shown by  \squared{1}. 
In  \squared{2}, node $j$, as a PN, reserves a time-slot for its transmission in the scheduled section if it has not already been chosen as a PN.
Then, in \squared{3} it broadcasts a message to inform its neighboring nodes about the reserved time-slot in the scheduled section as well as other  information which is required for the algorithm to run.
The required information depends on the cost sharing scheme used in the network which we discuss in the next section. 
In the interval \squared{4}, the CN $i$ informs its neighboring nodes about its new PN.
Finally, \squared{5} is a guard interval which can also be used for other nodes to process the information they receive in \squared{3} and \squared{4} from the nodes $i$ and $j$.

In our model, a node accesses the channel at most once in a RACH phase.
The random access section is prone to collision, and a collision occurs if two or more nodes use \textit{the same time-slot} in the RACH phase for sending their requests.
In this case, the nodes have to retry accessing the channel in the next round of RACH phase.
The collision probability depends on the number of nodes, the number of slots in a RACH phase and the frequency of access. 
By a proper design of the MAC mechanism, one can keep the average number of nodes which face a collision significantly low \cite{Mahdi_thesis}.
It is important to remark that in this paper, we do not focus on random channel access optimization.
Moreover, to have a fair comparison with the existing works, through the rest of this paper we abstract from the collision probability  and measure the performance of our new scheme separately.
We  assume that the possible collisions  impact the performance of our work and the existing works in a comparable way.
Indeed, in this work, given a random channel access method for multi-hop networks, we propose  a decentralized algorithm that finds an energy-efficient BT for data dissemination during the scheduled access section.

Since such a channel access scheme requires time synchronization at the nodes, it is common to use the clock of the source as a reference clock.
The synchronization can be done   via  a dedicated time-slot  in a  hop by hop manner from the source toward the leaves of the BT \cite{Wu_SPmag_11}.
We assume that synchronization in the network is attained.
It should also be noted that, although the signaling messages impose additional energy consumption on the network, we assume that the imposed energy is negligible compared to that required for the actual data dissemination.
We further discuss the overhead issue in the next section.

Due to the transmit power constraint, a node $j$ can be a PN of node $i$ if the power required for the link between the nodes $j$ and $i$ is less than $p_j^\mathrm{max}$.
The set of  neighboring nodes of node $i$ is denoted by $\mathcal{N}_i$ and defined as 
\begin{equation}
\label{eq:neigbdef}
\mathcal{N}_i = \left\{ j \bigg\arrowvert j \in \mathcal{Q} , p_{i,j}^\mathrm{req} \leq p_j^\mathrm{max} \right\}.
\end{equation}
It is assumed that every node knows the channel gains of the links to its neighboring nodes.
We specifically denote the unicast transmit power required for the link between node $j$ and its neighboring node $i$ by $p_{i,j}^\mathrm{uni}$.
In other words, $p_{i,j}^\mathrm{uni} = p_{i,j}^\mathrm{req}$, if $j \in \mathcal{N}_i$, see Fig. \ref{fig:net}.
Considering the circuitry power of PN $j$, the total power required for a unicast transmission at PN $j$ is 
\begin{equation}
\label{eq:powtx_c_uni}
P_{i,j}^\mathrm{uni} = p_j^\mathrm{c} + p_{i,j}^\mathrm{uni} .
\end{equation}
In case of multicast transmission, where a  parent node $j$ has multiple CNs, the total required power at a PN $j$ in \eqref{eq:node_pow_def} is given by
\begin{equation}
\label{eq:powjdef2}
P_j (\mathbf{a} ) = P_j (\mathcal{M}_j) =  
\begin{cases}
\begin{aligned}
\max_{i \in \mathcal{M}_j} & \left\{  P_{i,j}^\mathrm{uni} \right\} & \quad &\text{if } \mathcal{M}_j \neq \varnothing \\
0& & \quad & \text{if } \mathcal{M}_j = \varnothing \\
\end{aligned}
\end{cases}
.
\end{equation}

Finally, the \textit{total} required power in the \textit{network} for message dissemination among all the nodes, simply termed the \textit{network power}, is   calculated by
\begin{equation}
\label{eq:ptot}
P^\mathrm{net} (\mathbf{a}) = \sum_{ j \in \mathcal{Q}}  P_j (\mathbf{a} ).
\end{equation}

It should be remarked that  the message flow from the source to the nodes must result in a tree-graph, rooted at the source without any cycle.
When a cycle occurs in a graph, a part of the network loses its connections to $\mathrm{S}$.
We define the route of a node as the set of the nodes which are on the route from $\mathrm{S}$ to node $i$, including node $i$, and denote it by  $\mathcal{R}_i$.
For instance,  $\mathcal{R}_w = \{\mathrm{S},k,u,w\}$ for the given BT in Fig. \ref{fig:net}.
The route of $\mathrm{S}$ is set to $\mathcal{R}_\mathrm{S}=\{\mathrm{S}\}$. If node $i$ chooses PN $j$, $\mathcal{R}_i$ can be simply found as
\begin{equation}
\label{eq:route-find}
\mathcal{R}_i = \mathcal{R}_j \cup \{i\}.
\end{equation}
Note that  using interval \squared{4} in Fig. \ref{fig:chaccess}, node $i$ informs its neighboring nodes about its new route.
The network-wide objective, which is also referred  to as the \textit{global objective}, is to minimize the network power defined in \eqref{eq:ptot} such that every receiving node in $\mathcal{W}$ receives the source's message from a node $j \in \mathcal{Q} \backslash \{i\}$ and has the source in its route as
\begin{IEEEeqnarray}{lll} \label{eq:gob}
 \underset{ \{\mathcal{M}_j\}_{ \forall j\in \mathcal{Q} } } {\text{minimize}} \quad & \sum_{ j \in \mathcal{Q}} P_j (\mathcal{M}_j) &\\
\text{subject to:}   &   \forall i \in \mathcal{W} \quad \exists! \text{ } j \in \mathcal{Q} \backslash \{i\}  :  i \in \mathcal{M}_j, \mathrm{S} \in \mathcal{R}_j \quad \notag  
\end{IEEEeqnarray} 
Since every node $i \in \mathcal{W}$ is allowed to choose one PN and the source must be in the route of the PN, i.e.,  $\mathrm{S} \in \mathcal{R}_j$, the constraints above give us a tree-graph.

\section{Game-Theoretic Algorithm} \label{sec:game}
\subsection{Game design and properties} \label{sec:game:a}
The game is characterized by the set $\mathcal{W}$ of non-cooperative and rational nodes, that is, all the nodes in the network except the source.
The proposed game is a  dynamic (iterative) game such that at every iteration $t$, one of the nodes of the network takes one of its possible actions from its action set.  
 The action of a node $i \in \mathcal{W}$ in this game is to choose another node $j \in \mathcal{Q} \backslash \{i\}$ as its PN to connect to and receive the source's message from.
We  denote  the action of node $i$ as $a_i \in \mathcal{A}_i^{(t)}$ in which $\mathcal{A}_i^{(t)}$ is the action set of player $i$ at iteration  $t$. 
The action profile of the game is shown by $\mathbf{a} = (a_1, \dots, a_N) \in \mathcal{A} ^{(t)}$ in which $\mathcal{A} ^{(t)}= \mathcal{A}_1^{(t)} \times \dots \times \mathcal{A}_N^{(t)} $ is the joint action set of the game at iteration $t$.
The action profile of the game can also be  denoted by $\mathbf{a} = (a_i, \mathbf{a}_{-i})$ in which $\mathbf{a}_{-i}$  represents  the actions of all the players except   node $i$.
Further, $\mathbf{a} \overset{i}{\rightarrow} \mathbf{a}'$  denotes an update in the action profile of the game from $\mathbf{a}$ to $\mathbf{a}' = (a_i', \mathbf{a}_{-i})$ when node $i$ updates its action from $a_i$ to $a_i'$.
The total power required in the network depends on the action profile of the game, i.e., which nodes are chosen as PNs.
 
We denote the action profile corresponding to the optimum BT by $\mathbf{a}^\mathrm{opt}$.
Based on the action profile of the game at iteration $t$, i.e., $\mathbf{a} \in \mathcal{A}^{(t)}$, a non-negative cost is assigned to every player of the game.
The cost function is defined as $C_i (a_i, \mathbf{a}_{-i}): \mathcal{A}^{(t)} \rightarrow \mathbb{R}^+ \cup \{0\}, \forall i \in \mathcal{W}$ in which $\mathbb{R}^+$ represents the positive real numbers.
We show the cost of node $i$ in case of choosing PN $j$ as $C_i(j,\mathcal{M}_j)$ since the cost just depends on the set of the nodes who choose the same PN.
The non-cooperative dynamic game $G$ is   defined formally by the tuple $G := < \mathcal{W}, \{\mathcal{A}_i\}_{i \in \mathcal{W}}, \{C_i\}_{i \in \mathcal{W}}>$.

The game $ G $ is a child-driven game, that is,  a node  as a  child  selects  another node in its neighboring area as its PN. 
The action set of a node has to be defined in a way to ensure that no cycle occurs in different iterations of the game.
Based on  the definition,  a cycle occurs in a rooted tree  when a node $i \in \mathcal{W}$ connects to one of its descendants  \cite{diestel2006};  
The descendants of a node $j\in \mathcal{Q}$ are all the nodes who have the node $j$ on their route to $\mathrm{S}$ and  a cycle occurs if it chooses one of its  descendants as its PN.
For instance in Fig. \ref{fig:net}, if node $k$ chooses node $w$ as its PN.
Denoting the route of a node $j$ at iteration $t$ by $\mathcal{R}_j ^{(t)}$,  we define the action set of a node $i \in \mathcal{W}$ at iteration $t$  as all the neighboring nodes of node $i$ except its descendants as
 \begin{equation}
\label{eq:ActionSet}
\mathcal{A}_i^{(t)} = \left\{ j \bigg\arrowvert   j \in \mathcal{N}_i, \mathrm{S}\in \mathcal{R}_j ^{(t-1)}  , i \notin \mathcal{R}_j ^{(t-1)}  \right\} 
 \end{equation}
 in which $\mathrm{S}\in \mathcal{R}_j ^{(t-1)}$  indicates node $j$ in order to be a PN of node $i$ must be  connected to the BT. 
For simplicity, we omit the time indicator $t$ in the $\mathcal{A}^{(t)} $.

In order to benefit from the broadcast nature of the wireless channel, the cost of the nodes should be defined in a way to motivate the CNs to form a multicast group and choose a common PN.
Moreover, the circuitry power of a transmitting node must also be considered in the cost model.
  The cost function in this game, based on the  MC  principle \cite{MAS08},  is defined  as 
\begin{IEEEeqnarray} {rl} \label{eq:MCdef}
C_i^\mathrm{MC}(j, \mathcal{M}_j)  = &P_j  (\mathcal{M}_j) - P_j  (\mathcal{M}_j  \backslash \{i\}), \qquad i \in \mathcal{M}_j 
\end{IEEEeqnarray}
in which $\mathcal{M}_j  \backslash \{i\}$ represents the set of  CNs of PN $j$ except the CN $i$.
Roughly speaking, the cost of node $i$ is the difference in the total power required at node $j$ with and without node $i$.
Based on  \eqref{eq:MCdef}, a positive cost is assigned to the CN that requires the highest unicast transmit power from  PN $j$ while  the cost assigned to  other CNs in $\mathcal{M}_j$  is zero.
The game $G$ with the MC, defined in \eqref{eq:MCdef}, as its cost function is called the CSG-MC.

To illustrate the cost model in \eqref{eq:MCdef}, let us assume that node $i$ and node $l$ require the highest and the second highest unicast powers form  PN $j$, respectively, see Fig. \ref{fig:net}.
In this case, the cost assigned to the CN $i$ using \eqref{eq:MCdef} is given by
\begin{IEEEeqnarray} {rl}
 \label{eq:ci_case1_j_tozih}
C_i^\mathrm{MC}(j,\mathcal{M}_j) &=   p_j^\mathrm{c} +    p_{i,j}^\mathrm{uni}  - \left( p_j^\mathrm{c}  +     \max_{h \in \mathcal{M}_j \backslash \{i\} } \{ p_{h,j}^\mathrm{uni}  \} \right) = p_{i,j}^\mathrm{uni} - p_{l,j}^\mathrm{uni} . 
\end{IEEEeqnarray}
In this case, either $i \in \mathcal{M}_j$ or $i \notin \mathcal{M}_j$,  the circuitry power is required at PN $j$ as it must serve the CN $l$.
Therefore, no \textit{additional} power, here the circuitry power,  is imposed on PN $j$ by CN $i$  and hence, the circuitry power of PN $j$ does not appear in the cost assigned to the CN $i$.
Moreover, if we assume that the CN $i$ is the only CN of the PN $j$, then based on \eqref{eq:MCdef}, the cost of CN $i$ contains the circuitry power of PN $j$, i.e.,  $C_i^\mathrm{MC}(j,\mathcal{M}_j) = p_j^\mathrm{c} + p_{i,j}^\mathrm{uni}$, as $\max_{h \in \mathcal{M}_j \backslash \{i\} } \{ p_{h,j}^\mathrm{uni} \} = 0$.
In other words,  since in a unicast both transmit and circuitry powers are imposed  on the PN $j$ by the CN $i$, the  circuitry power appears in the cost assigned to the CN $i$ as well as the   transmit power.

Therefore, depending on the structure of the BT and the transmission scheme (unicast or multicast), the cost model in \eqref{eq:MCdef} keeps or removes the circuitry power from the cost of receiving nodes to prevent establishing a new unicast transmission or   motivate the nodes to form a multicast receiving group, respectively.
Whether joining a multicast group is better than establishing a unicast is decided by the node based on its cost function. 
 
We employ the best response dynamics (BRD) for game $G$ such that at every iteration of the game, one of the players chooses an action as its best response to the action of other players.
The best response of player $i$, which is also referred  to as the \textit{local objective}, is defined as
\begin{equation}
\label{eq:lob}
a_i = \argmin_{j \in \mathcal{A}_i } \quad C_i(j, \mathcal{M}_j), \qquad \forall i \in \mathcal{W}, j \in \mathcal{N}_i.
\end{equation}
 
Finally, we consider an NE as the converging point of the state of the BT.
\begin{definition} \textit{\textbf{ (NE)}} 
An action profile $\mathbf{a}^* \in \mathcal{A}$ is an (pure) NE of the game $G$ if  
\begin{equation}
\label{eq:NEdef}
C_i(a_i^*,\mathbf{a}^*_{-i}) \leq C_i(a_i,\mathbf{a}^*_{-i}), \quad  \forall i \in \mathcal{W}, a_i \in \mathcal{A}_i.
\end{equation}
\end{definition}
While with the BRD, only one node is allowed to update its action in a time-slot, it is possible to enable multiple simultaneous actions per time-slot.
In \cite{ZhangPrasad_JSAC11}, a randomized distributed algorithm is introduced by which the game can be viewed as a time homogeneous Markov Chain with finite number of states in which each NE represents an absorbing state.
This will ensure convergence to an absorbing state with probability 1, even with simultaneous updates.
Although by such a design the game converges to an NE,  the complexity of the game increases, especially for designing rules for preventing the occurrence of  cycles in the BT.
 
\subsection{ Convergence and Discussion}\label{subsec:discuss} 
In this subsection, we discuss the properties of the game described in the previous subsection.
We first show that the game converges to an NE.
Then we discuss the properties of the game.

\begin{definition} \label{def:exct_algn}
Let 
$\underset{\mathbf{a} \rightarrow \mathbf{a}'}{\Delta^i} f :=  f(\mathbf{a}' ) - f(\mathbf{a} )$ be the change in the function $f$ when $\mathbf{a} \overset{i}{\rightarrow} \mathbf{a}'$.
The local objective in \eqref{eq:lob} is said to be aligned with the global objective in \eqref{eq:gob} if for every $i, \mathcal{W}, \underset{\mathbf{a} \rightarrow \mathbf{a}'}{\Delta^i} C_i <0$, then, $\underset{\mathbf{a} \rightarrow \mathbf{a}'}{\Delta^i} P^\mathrm{net} <0$.
They are also said to be exactly aligned if $\underset{\mathbf{a} \rightarrow \mathbf{a}'}{\Delta^i} C_i = \underset{\mathbf{a} \rightarrow \mathbf{a}'}{\Delta^i} P^\mathrm{net}$.
\end{definition}
 
\begin{definition} \textit{\textbf{(Potential game) }} 
\label{def:pot}
A game $G$ is an exact potential game \cite{Monderer96} if there exists a function $\Phi: \mathcal{A} \rightarrow \mathbb{R}$, called the potential function, such that $\forall i \in \mathcal{W}$,
$ \mathbf{a}, \mathbf{a}' \in \mathcal{A}$
\begin{equation} \label{eq:PotDef}
\underset{\mathbf{a} \rightarrow \mathbf{a}'}{\Delta^i} C_i = \underset{\mathbf{a} \rightarrow \mathbf{a}'}{\Delta^i} \Phi
\end{equation}
\end{definition}

\begin{theorem} \label{th:potthm}
The game $G$ with the proposed MC  cost sharing scheme is an exact potential game with the potential function 
\begin{equation}
\label{eq:potfun_defth}
\Phi (\mathbf{a}) = \sum _{j \in \mathcal{Q}}  P_j (\mathbf{a}).
\end{equation}
\end{theorem}
\begin{IEEEproof}
We verify  \eqref{eq:PotDef} with the cost function and the potential function, introduced in $\eqref{eq:MCdef}$ and \eqref{eq:potfun_defth}, respectively.
Let us assume that  $a_i = j$ and  $a_i' = k$ and 
$\mathbf{a} \overset{i}{\rightarrow} \mathbf{a}'$, 
see Fig.  \ref{fig:net}.
With such a transition, just PN $j$ and PN $k$ will be affected among  the PNs in the network.
Thus, the network power, here the potential function of the game,  can be written as
\begin{equation}
\label{eq:potfun2nd}
\Phi(\mathbf{a}) = \sum _{j \in \mathcal{Q}}  P_j (\mathbf{a}) = P_j (\mathcal{M}_j) + P_k(\mathcal{M}_k) + \sum_{  \mathclap{m \in \mathcal{Q}\backslash \{j,k\} }} P_m(\mathcal{M}_m) .
\end{equation}
The cost of node $i$ when $i \in \mathcal{M}_j$ is given by
\begin{equation}  
 \label{eq:ci_case1_j}
C_i^\mathrm{MC}(j,\mathcal{M}_j)  =   P_{j} (\mathcal{M}_j ) - P_{j} (\mathcal{M}_j \backslash \{i\}),  
\end{equation}
and the cost assigned to node $i$ when it joins PN $k$ is
\begin{equation}
\label{eq:ci_case1_k}
C_i^\mathrm{MC}(k, \mathcal{M}_k \cup \{i\}) = P_k (\mathcal{M}_k \cup \{i\} ) - P_k (\mathcal{M}_k).  
\end{equation}
The potential function in \eqref{eq:potfun2nd} when  $a_i' =k$ is given by
\begin{IEEEeqnarray} {rl}
\label{eq:pot_th1_ik}
\Phi( \mathbf{a}')   = P_{j} (\mathcal{M}_j \backslash \{i\} ) + P_{k} (\mathcal{M}_k \cup \{i\}) + \sum_{  \mathclap{m \in \mathcal{Q}\backslash \{j,k\} }}  P_m(\mathcal{M}_m).
\end{IEEEeqnarray}
Then, using (\ref{eq:potfun2nd}-\ref{eq:ci_case1_j}-\ref{eq:ci_case1_k}-\ref{eq:pot_th1_ik}) we have
\begin{IEEEeqnarray} {rl}
\label{eq:pot_th1_fin}
\underset{\mathbf{a} \rightarrow \mathbf{a}'}{\Delta^i} \Phi =
P_{k} (\mathcal{M}_k \cup \{i\}) - P_{k} (\mathcal{M}_k)- P_j(\mathcal{M}_j ) + P_{j} (\mathcal{M}_j \backslash \{i\} )
=
\underset{\mathbf{a} \rightarrow \mathbf{a}'}{\Delta^i} C_i^\mathrm{MC}
 \end{IEEEeqnarray}
\end{IEEEproof}
 \begin{corollary} \label{col:NEtheBest}
 If  the cost of the nodes is defined based on the MC, the local objective in the game $G$ is exactly aligned with the global objective defined  in \eqref{eq:gob}. 
\end{corollary}
\begin{corollary} \label{col:monotone}
The BRD converges to an (a pure) NE for the game $G$.
\end{corollary}

\begin{IEEEproof}
Since the game $ G$ is an exact potential game, it possesses a pure NE \cite{Monderer96}. 
An NE of the game is any action profile $\mathbf{a}^*$ that (locally) minimizes the potential function in \eqref{eq:potfun_defth}.
When a node updates its action in order to reduce its cost, based on Theorem \ref{th:potthm}  the same reduction occurs in $\Phi$.
As  $\Phi$,  i.e., the   network power, is bounded from below, after some iterations the game $G$ reaches a state at which none of the nodes can further reduce its own cost.
\end{IEEEproof}

\begin{remark}\label{rmk:complexity}
Although reaching an NE in a finite number of iterations is guaranteed by using BRD, its convergence rate
is exponential in the worst case.
Nevertheless, the average convergence rate of the BRD is $O(N)$ \cite{Durand_agt_springer_16} which is acceptable for practical scenarios.
\end{remark}

\begin{theorem}
\label{th:optNE}
Using  MC, $\mathbf{a}^\mathrm{opt}$ is always an NE of the game $G$.
\end{theorem}

\begin{IEEEproof}
Recall that $\mathbf{a}^\mathrm{opt}$ is the action profile of the game associated with the optimum BT.
Let us assume  that the $\mathbf{a}^\mathrm{opt}$ is not an NE.
Therefore, based on the definition,  at least one of the nodes of the network can update  its action to reach  a lower cost.
As shown in Theorem \ref{th:potthm}, reduction in the cost of a node results in the same reduction of $\Phi$, that is, the  network power.
This is a contradiction as the BT of $\mathbf{a}^\mathrm{opt}$ is optimum.
\end{IEEEproof}

In a CSG with MC as the sharing scheme, the aggregated cost paid by  CNs is not necessarily equal to the total power required at their corresponding PN.
This property makes the MC a non-budget-balanced scheme.
We now discuss the properties of ES and SV schemes if one applies them to the MPBT problem.
ES and SV are two of the widely-adopted budget-balanced schemes in the field of  CSGs and are known to  be fair depending on the application \cite{Anshelevich:2004:faircost, Mousavi_ICC16}.

\begin{definition} \textbf{(Budget-balanced cost sharing scheme \cite{MAS08} )} \label{def:BB}
A cost sharing scheme $C^\mathrm{BB}(.)$ is budget-balanced  if for any node $j \in Q$  
\begin{equation}
 \label{eq:BB-def} 
\sum_{i \in \mathcal{M}_j} C_i^\mathrm{BB}(j,\mathcal{M}_j) = P_j (\mathcal{M}_j). 
\end{equation}
\end{definition}

\begin{definition} \textbf{(The Shapley value (SV) \cite{S53})} 
 
Let  $p_{0,j}^\mathrm{uni} = 0 \leq p_{1,j}^\mathrm{uni}\leq  \dots \leq p_{i,j}^\mathrm{uni} \leq \dots \leq p_{|\mathcal{M}_j|,j}^\mathrm{uni}$, be the sorted individual unicast powers imposed on PN $j$ by its CNs, then the Shapley value  of the $i$-th CN  is given by \cite{LO73}

 \begin{equation}
 \label{eq:SHAP-def} 
C_i^\mathrm{SV}(j,\mathcal{M}_j) = \sum\limits_{n=1}^{i}\frac{P_{n,j}^{\mathrm{uni}} - P_{n-1,j}^{\mathrm{uni}}}{|\mathcal{M}_j|+1-n}.
\end{equation}
\end{definition}

\begin{definition} \textbf{(Equal-share (ES) cost sharing scheme)} 
A cost sharing scheme is ES if the cost is shared among the CNs of a PN equally, regardless of the individual unicast powers  required for the links to the CNs  \cite{MAS08}, i.e.,
\begin{equation}
 \label{eq:EQS-def} 
C_i^\mathrm{ES}(j,\mathcal{M}_j) = P_j (\mathcal{M}_j)/|\mathcal{M}_j|, \quad \forall i \in \mathcal{M}_j.
\end{equation}
\end{definition}

 \input{theorem_bb_v1}

\begin{figure*} [!t]
	\begin{minipage}{.45\textwidth}
 		  \centering
 		  \includegraphics[width=.9 \columnwidth]{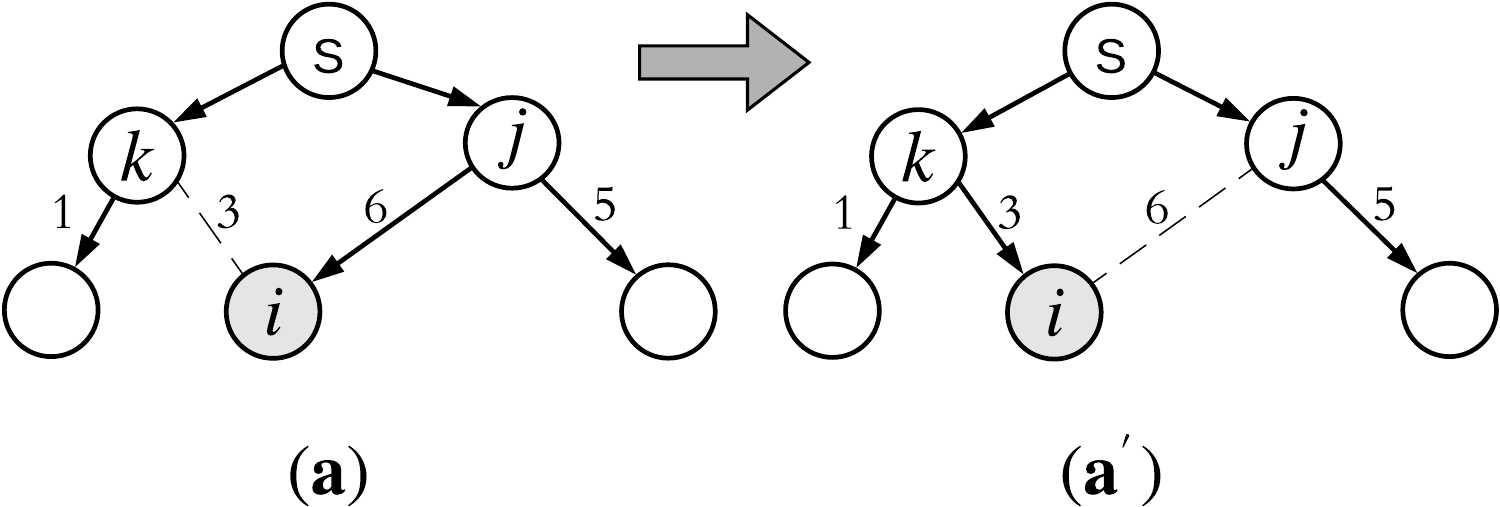}
 		  \caption{With ES and SV the optimum BT is not guaranteed to be an NE of the game.}
 		  \label{fig:global-local}
	\end{minipage} \qquad
	\begin{minipage}{.45\textwidth}
			\centering
   			\includegraphics[width=.9 \columnwidth]{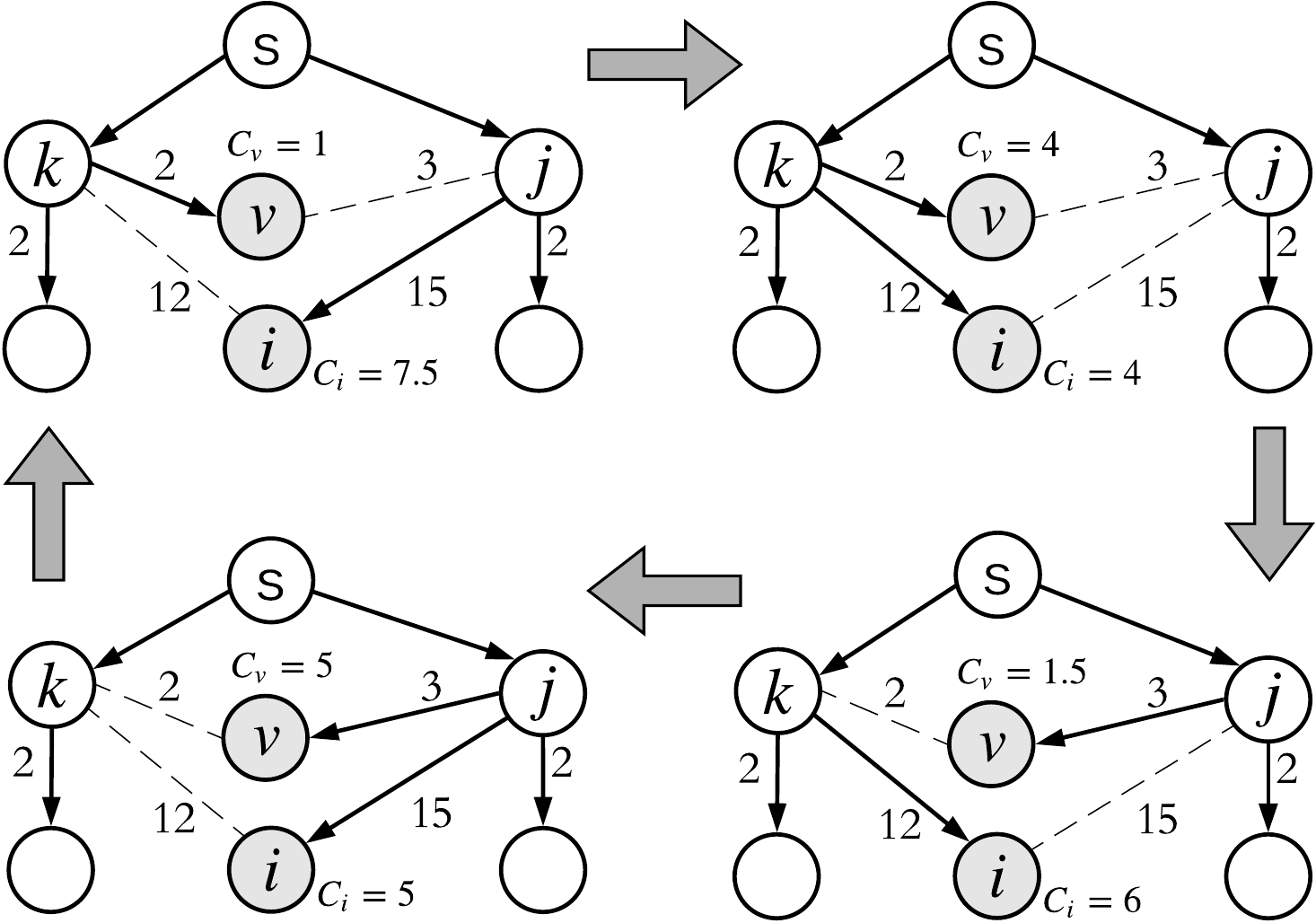}
   			\caption{The ES  scheme for the MPBT problem does not guarantee the convergence.   }
   			\label{fig:ES-problem}
   	 \end{minipage} 
\end{figure*}

In the rest, we further investigate the properties of ES and SV in comparison to MC.
Before that we define the following property for a cost function.

\begin{definition} \label{def:cross} \textbf{(cross-monotonicity)} 
A set function $f: 2^\Omega \rightarrow \mathbb{R} $ is cross-monotone if for every $\mathcal{S} \subset \Omega$, $i \in \mathcal{S}$ and   $k \in \Omega\backslash \mathcal{S}$, $f(\mathcal{S} \cup \{k\}) \leq f(\mathcal{S})$ \cite{MAS08}.
\end{definition}
According to Definition~\ref{def:cross}, a cost function is cross-monotone if the cost of the CNs of a PN does not increase when a new node joins the PN.
\begin{lemma} \label{lem:cross-mono}
A necessary condition of a budget-balanced cost sharing scheme  to guarantee the existence of an NE is cross-monotonicity \cite{Marden_TAutC13}.
\end{lemma}

\begin{theorem} \label{th:ES-MPBT}
The ES does not guarantee the existence of an NE for the MPBT problem.
\end{theorem}
\begin{IEEEproof}
It is easy to see that based on Definition \ref{def:cross}, the ES is not cross-monotone and hence, based on  Lemma \ref{lem:cross-mono} the convergence to an NE is not guaranteed.
Moreover, we provide an instance of the network in Fig. \ref{fig:ES-problem}  for which the ES  scheme does not lead to an NE.
In this figure, updating the action at node $i$  increases the cost of node $v$ and vice versa. 
Hence, the nodes $i$ and $v$ iteratively update their actions and the game $G$ does not converge.
\end{IEEEproof}

\begin{lemma} \label{lem:shapley-equal}
When the power a PN $j$ is fixed and independent of $\mathcal{M}_j $, the cost assigned to a CN by the SV is equal to the one assigned by the ES, i.e., $C_i^\mathrm{ES}(j, \mathcal{M}_j) = C_i^\mathrm{SV}(j, \mathcal{M}_j), \forall i \in \mathcal{M}_j$.
\end{lemma}
\begin{IEEEproof}
With fixed $P_j (\mathcal{M}_j)$, the contribution of every CN in $\mathcal{M}_j$ on the transmit power of PN $j$ is then equal and can be assumed as   $P_{i,j}^\mathrm{uni} = P_j (\mathcal{M}_j), \forall i \in \mathcal{M}_j$.
Using \eqref{eq:SHAP-def}  for $i=1$  we have $C_1^\mathrm{SV}(j,\mathcal{M}_j) = P_j (\mathcal{M}_j)/|\mathcal{M}_j|$. 
We assume $P_{i,j}^{\mathrm{uni}} = P_{i-1,j}^{\mathrm{uni}}, \forall i >1$, hence, $C_i^\mathrm{SV}(j,\mathcal{M}_j) = C_1^\mathrm{SV}(j,\mathcal{M}_j) ,\forall i \in  \mathcal{M}_j$.
\end{IEEEproof}
Note that  for the MTBT problem where the transmit powers are all equal and fixed (and their values do  not matter), the ES shares the cost as $C_i^\mathrm{ES}(j,\mathcal{M}_j) = 1/|\mathcal{M}_j|$.

\begin{lemma} \label{lem:shapley-cnvg}
A non-cooperative CSG  with SV scheme is a potential game for which an NE always exists  \cite{Monderer96} \cite{Gopalakrishnan:potgame:14}.
 \end{lemma}
 
\begin{theorem} \label{th:ES-MFPBT}
The ES guarantees the existence of an NE for the MFPBT (and MTBT) problem.
\end{theorem}
\begin{IEEEproof}
As stated in Lemma \ref{lem:shapley-equal}, the ES scheme is a  special case of the SV when the contributions of the CNs are assumed to be equal.
This is the case for the MFPBT problem where the transmit power of a PN, regardless of the individual unicast powers required for the links to its CNs, is fixed.
Hence, using the ES  for the MFPBT problem can be seen as a special case of  the MPBT problem with the SV scheme.
Since based on Lemma \ref{lem:shapley-cnvg}, the SV guarantees the existence of an NE for the MPBT problem, the ES does so for the MFPBT problem.
\end{IEEEproof}
Note that, based on Definition \ref{def:cross}, the ES  is cross-monotone for the MFPBT problem and fulfills the necessary condition in Lemma \ref{lem:cross-mono}.

In designing games for decentralized optimization, the elements of the game such as cost function, action sets and the strategy of the nodes have to be designed in a way to guarantee that the individual local behavior of the players is desirable from the global system point of view.
\begin{remark} \label{rem:info_req}
The implantation of different cost sharing schemes differs in terms of the information overhead they require.
The ES is the simplest one since  a node, to calculate its cost, just requires knowing  the number of CNs in a multicast receiving group.
With MC, every node needs to know the highest and the second highest unicast powers required by the CNs of a PN.
Finally, the SV imposes the highest overhead on the network.
To calculate the cost using the SV, a node must know the unicast power required  of every individual  CN in a multicast group.
\end{remark}
The information  required for decision making  has to be transmitted in a neighboring area by every node as overhead information via a broadcast channel.
Table \ref{tab:compare} summarizes the properties of the different cost sharing schemes. 
The comparison in terms of overhead is relative.
 
\begin{table} [!t]  
	\centering
	\footnotesize
  	\caption{ Properties of different cost sharing schemes } 
\begin{center}
\small
 \begin{tabular}{|c||c|c|c|} 
 \Xhline{2.2 \arrayrulewidth}
   & MC & SV & ES  \\
 \hline
 \Xhline{2.2 \arrayrulewidth}  
  \addlinespace[.5mm]
 \Xhline{2.2 \arrayrulewidth}  
Convergence for MPBT  &  yes & yes & no  \\ \hline
Convergence for MTBT/MFPBT &  yes & yes & yes \\   \hline
Is $\mathbf{a}^\mathrm{opt}$ always an NE?  & yes & no & no  \\ \hline
Overhead & medium  & high & low  \\
\hline 
\Xhline{2.2 \arrayrulewidth}
 \end{tabular}
\end{center}
\label{tab:compare}
\end{table}

In conclusion, based on what has been discussed and using Table \ref{tab:compare}, we can find that the MC has two main advantages over the SV for the MPBT problem.
Firstly, with MC, unlike SV, $\mathbf{a}^\mathrm{opt}$ is always an NE.
Secondly, the required overhead information for MC is lower than that of the SV.
This becomes more important when the size of the multicast receiving group increases.
Note that here we do not consider the ES for the MPBT problem due to the lack of convergence guarantee.
 
The efficiency of a game-theoretic scheme can be studied by analyzing the worst-case outcome for which the  measures of the price of anarchy (PoA) and the price of stability (PoS) are used.
\begin{definition} \label{def:poa_pos} \textbf{(PoA and PoS) } 
Let $\mathcal{E}($G$)$  be the set of NEs of the game $G$ and $\mathcal{G}$ denote the set of all possible games $G$.
The PoS and the PoA of the game $G$ are defined respectively as \cite{MAS08}
\begin{equation} \label{eq:posg}
\mathrm{PoS}(\mathcal{G}) :=  \sup_{G \in \mathcal{G}}  \frac{\min_{a\in \mathcal{E}(G)}  P^\mathrm{net} (\mathbf{a})}{ P^\mathrm{net} (\mathbf{a^\mathrm{opt}}) }, \quad \mathrm{and}  \quad 
\mathrm{PoA}(\mathcal{G}) :=  \sup_{G \in \mathcal{G}} \frac{\max_{a\in \mathcal{E}(G)}  P^\mathrm{net} (\mathbf{a})}{ P^\mathrm{net} (\mathbf{a^\mathrm{opt}}) }.
\end{equation}
\end{definition}
\begin{observation} 
The PoS of the proposed game is 1.
\end{observation}
\begin{IEEEproof}
Based on Theorem \ref{th:optNE}   $\mathbf{a}^\mathrm{opt}$ is always an NE of the game $G$, thus, $\mathrm{PoS}(\mathcal{G}) = 1$.
\end{IEEEproof}

\begin{theorem}
\label{th:poa}
 The PoA of the game $G$ is lower bounded by $\Omega(N^{(\alpha -1)})$.  
\end{theorem}
 \begin{IEEEproof}
 The proof has been provided in Appendix \ref{app:poa}.
 \end{IEEEproof}
 
Although the lower bound on the PoA provides an insight into the performance of the game, it is of limited utility.
Finding an upper bound for the PoA is not straightforward due to the randomness in the circuitry power $p^c$ of the nodes as well as their maximum transmit power $p^\mathrm{max}$.
Although one can let $p^c$ and/or $p^\mathrm{max}$ be equal for all the users, finding  the PoA is still challenging due to the geometrical constraints of the problem.
More precisely,  the unicast power required for the link between two nodes cannot be any arbitrary number and depends on the distance of the nodes.
Such a constraint makes the calculation of the PoA complicated and we leave it as an interesting open problem.

\section{Centralized Approach} \label{sec:MILP}
 
In this section, we model the MPBT problem of \eqref{eq:gob} as an MILP.
The proposed MILP is inspired by the one proposed by the authors of \cite{Chen_TMC13} for the MTBT problem where the transmit power control is omitted and the circuitry power is not considered.  
The provided MILP in \cite{Chen_TMC13} mainly finds the optimum value of the network power by finding the nodes that should act as transmitting nodes as well as their transmit power.
It does not determine the structure of the optimum BT.
Hence, besides the MILP, we  propose an algorithm by which the structure of the optimum BT can be found based on the solution of the MILP. 
Before providing the MILP formulation for the MPBT problem, we define the following vectors and variables and, later in this section, explain them by a toy example:  
\begin{itemize}

\item 
 Transmission vector: the transmission vector is used to determine whether a node $j \in \mathcal{Q}$ acts as a transmitting node or not. Moreover, in case that node $j$ is a transmitting node, it determines the CN of PN $j$ that requires the highest unicast power. The transmission vector is defined as  $\mathbf{t}_j = [t_{1,j},  \dots, t_{N,j}]^\mathrm{T}, j \in \mathcal{Q}$  as an $N \times 1$  vector such that $t_{i,j} \in \{0,1\}$ and $t_{i,j} = 1$ if and only if node $i$ is the CN of PN $j$ that requires the highest unicast power among all nodes in $\mathcal{M}_j$.
Moreover, $\| \mathbf{t}_j\| \leq 1$ in which $\|.\|$ represents the norm operator. 
If node $j$ is a transmitter, then $\| \mathbf{t}_j\| =1$, otherwise $\| \mathbf{t}_j\| =0$.  

\item Reachability vector: it determines that if a node $i \in  \mathcal{W}$ is a CN of PN $j$ with highest required unicast power, given $P_j^\mathrm{Tx} = p_{i,j}^\mathrm{uni}$, which of the other nodes in $\mathcal{W}$ fall inside the coverage area of PN $j$ (without imposing additional transmit power on PN $j$). It is defined as $\mathbf{r}_{i,j} = [ r_{i,j}^{(1)}, \dots, r_{i,j}^{(l)}, \dots, r_{i,j}^{(N)}] $ as a $1 \times N $ binary vector for all $j \in \mathcal{Q}$ and $ i,l \in \mathcal{W}$.
The $l$-th entry of $\mathbf{r}_{i,j}$ is equal to 1 if $p_{l,j}^\mathrm{uni} \leq p_{i,j}^\mathrm{uni}  = p_j^\mathrm{Tx}, \forall i \in \mathcal{W}$. 
Since a node does not transmit to itself, then, $r_{i,i}^{(l)} = 0, \forall i,l \in \mathcal{W},r_{i,j}^{(j)} = 0, \forall i,j \in \mathcal{W}$.
Reachability matrix $\mathbf{R}_j = [ \mathbf{r}_{1,j}^\mathrm{T}, \dots, \mathbf{r}_{N,j}^\mathrm{T}]^\mathrm{T}, j \in \mathcal{Q} $ is an  $ N \times N$ binary matrix with $\mathbf{r}_{i,j}$ the reachability vector.
\item Downstream value: the downstream value $d_{i,j}$  is defined for the link between any two nodes $j$ and $i$ in $\mathcal{Q}$ and shows the total number of nodes in the network that rely on the transmission from PN $j$ to CN $i$ for receiving the source's message.
\end{itemize}
 
Since the outcome of the MILP must be a tree graph rooted at the source, i.e., $\mathrm{S} \in \mathcal{R}_i. \forall i \in \mathcal{W}$, it has been shown in \cite{Chen_TMC13} that three conditions for the downstream have to be met.
Firstly, the source cannot be in the downstream of any  other node as it is not a CN for other PNs. 
Secondly, the number of downstream nodes of the source node must be equal to $N$, as the whole network is connected to the source, either directly or indirectly.
Finally, the difference between the sum of the downstream values of the links coming in and going out of a node in $\mathcal{W}$ must   equal to 1.

\begin{figure}[t]
   \centering
   \includegraphics[width=.35 \columnwidth]{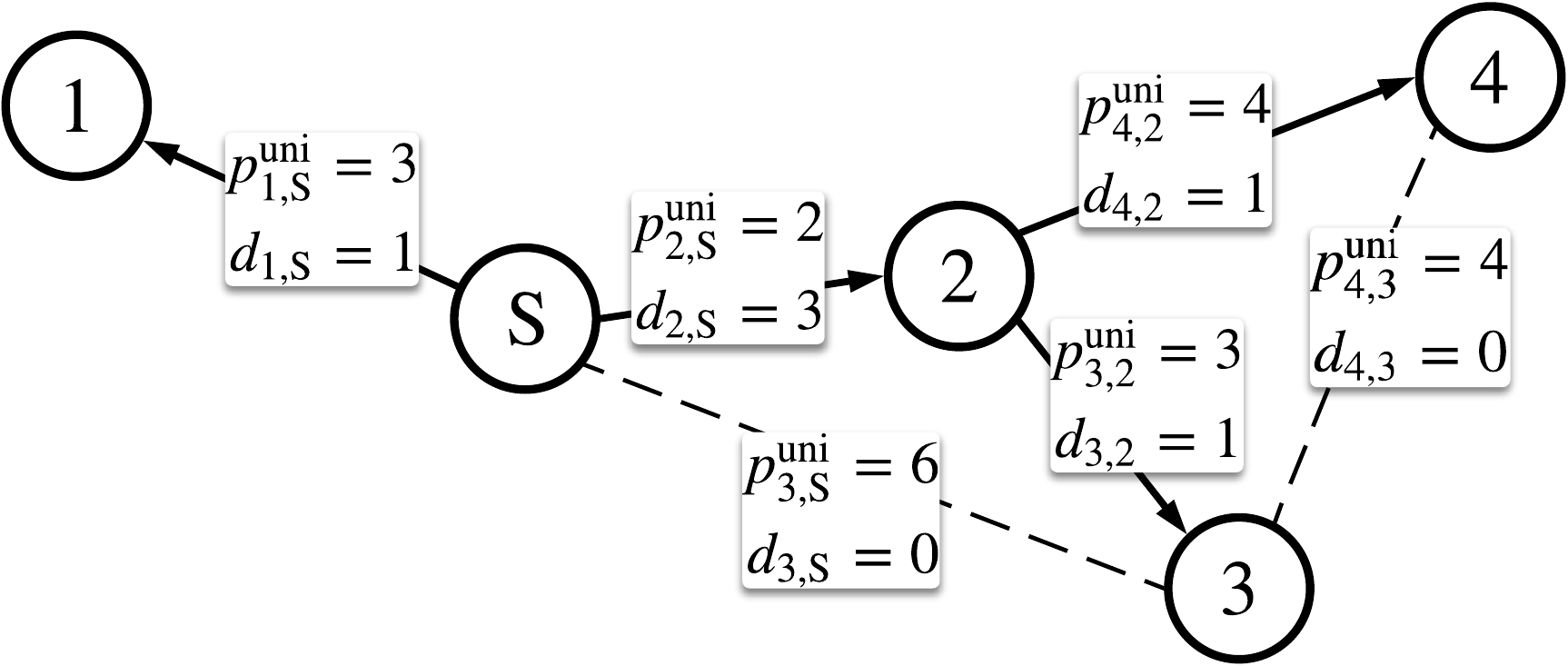}
   \caption{A sample BT from S to four other nodes. The numbers on the links  show  the required unicast powers, $p_{i,j}^\mathrm{uni}$, and the downstream values, $d_{i,j}$.}
   \label{fig:MILP}
\end{figure}

We explain the defined vectors and matrices in detail using the illustration shown in Fig. \ref{fig:MILP}.
In the BT of Fig. \ref{fig:MILP}, the source node  multicasts the message to  node  1 and node 2. 
Then, node 2 forwards the message to nodes 3 and 4.
$p_{i,j}^\mathrm{uni}$ required between any two nodes $i$ and $j$ and $d_{i,j}$ of the link are also shown in Fig. \ref{fig:MILP}.
As   can be seen, the downstream values of the links between the source and nodes 1 and 2 are equal to $d_{1,\mathrm{S}} = 1$ and $d_{2,\mathrm{S}} = 3$, respectively. 
Therefore, $d_{1,\mathrm{S}}+d_{2,\mathrm{S}} = 4$, that is, the total number of downstream nodes of  $\mathrm{S}$ is equal to the  total number of receiving nodes in $\mathcal{W}$.
The difference between the downstream values coming to and going out of node 2, as an intermediate node, is 1, i.e., $d_{2,\mathrm{S}} - \left( d_{3,2}+d_{4,2} \right) =1$.
This is also true for the nodes which do  not forward the message.\\
Variables  $t_{i,j}$ and $d_{i,j}$ are to be found by the MILP for all $i,j \in \mathcal{Q} $, while $\mathbf{R}_j$ can be obtained based on the unicast power required between the nodes.
Based on the unicast  power  for each link  shown in Fig. \ref{fig:MILP}, the reachability matrix for $\mathrm{S}$ is given by
\begin{equation} \label{eq:MILP_r}
\mathbf{R}_\mathrm{S} = 
\begin{bmatrix}
1&1&0&0\\
0&1&0&0\\
1&1&1&0\\
0&0&0&0
\end{bmatrix}.
\end{equation}
 The entries of the last row in \eqref{eq:MILP_r}, i.e., $\mathbf{r}_{4,\mathrm{S}}$, are all zero as node 4 and $\mathrm{S}$ are no neighbors, that is, node 4 cannot be reached by $\mathrm{S}$ due to the power constraint at $\mathrm{S}$.
It can be seen in \eqref{eq:MILP_r} that $\mathbf{r}_{1, \mathrm{S}} = [1,1,0,0]$, that is, $r_{1, \mathrm{S}}^{(1)}$ and $r_{1, \mathrm{S}}^{(2)}$ are equal to 1. 
Recall the entries of $\mathbf{r}_{i,j}$ show the nodes that can receive the message from PN $j$ without additional transmit power at node $j$ if the transmit power of node $j$ is equal to $p_{i,j}^\mathrm{uni}$.
This shows that if the source node transmits to node 1, then, node 2 can also receive the source's message by a multicast transmission without additional transmit power.  
In order to find which of the nodes of the network are covered by PN $j$ based on its transmission, we define the expression  $\mathbf{y}_j = [y_{1,j}, \dots, y_{N,j}]^\mathrm{T}$ in \eqref{eq:yj} as 
$\mathbf{y}_j = \mathbf{R}_j  ^\mathrm{T} \mathbf{t}_j$. 
More precisely, $\mathbf{y}_j$ is equal to one of the reachability vectors of node $j$ depending on  its transmission matrix $\mathbf{t}_j$. 
In the BT shown in Fig. \ref{fig:MILP}, $\mathbf{t}_\mathrm{S}=[1,0,0,0]^\mathrm{T}$.
Using \eqref{eq:MILP_r} and \eqref{eq:yj}, we have $\mathbf{y}_\mathrm{S}= \mathbf{r}_{1,\mathrm{S}}^\mathrm{T}=[1,1,0,0]^\mathrm{T}$.

The MILP for the MPBT problem is provided in Fig. \ref{fig:MILPeq}.
\begin{figure} 
\hrule 
\begin{subequations}
\begin{IEEEeqnarray}{llll} \label{eq:MILP}
 \underset{  t_{i,j}  } {\text{min}} \quad & & \sum_{ j\in \mathcal{Q}, i \in \mathcal{N}_j}  t_{i,j} P_{i,j}^\mathrm{uni} &\\
 \text{s.t. }   & && \notag \\
&&  \sum_{i \in \mathcal{N}_j} t_{i,j}= 
\begin{cases}
1  \\
\leq 1 
\end{cases}
&
\begin{tabular}{l}
$j$= $\mathrm{S}$\\
$j\in \mathcal{W}$
\end{tabular} \label{eq:const_a}  \\
&&  \sum_{i \in \mathcal{N}_j} \left( d_{i,j} -  d_{j,i}  \right) = 
\begin{cases}
N  \\
- 1 
\end{cases}
&
\begin{tabular}{l}
$j$= $\mathrm{S}$\\
$j\in \mathcal{W}$
\end{tabular} \label{eq:const_b}   \\
&&  \mathbf{y}_j = \mathbf{R}_j  ^\mathrm{T} \mathbf{t}_j  & \forall j \in \mathcal{W}   \label{eq:yj}  \\
&& d_{i,j} \leq N  y_{i,j}  & \forall j \in \mathcal{Q}, \forall i \in \mathcal{N}_j   \label{eq:const_d} \\
&& t_{j,j} = 0  & \forall j \in \mathcal{W}   \label{eq:const_e} \\
 & &  t_{i,j} \in \{0,1\},   d_{i,j}\geq 0, d_{i,j}\in \mathbb{R} \quad &  \forall i \in \mathcal{W}, j \in \mathcal{Q}. \label{eq:const_last}  
\end{IEEEeqnarray}
\end{subequations}
\hrule
\caption{The MILP formulation for MPBT problem.}
\label{fig:MILPeq}
\end{figure}
$P_{i,j}^\mathrm{uni}$ in \eqref{eq:MILP} is defined in \eqref{eq:powtx_c_uni}.  
Eq. \eqref{eq:const_a} expresses that the source node must be a transmitter while the other nodes $j \in \mathcal{W}$ are not necessarily   a transmitter. 
The constraints in \eqref{eq:const_b} and \eqref{eq:const_d}, as stated before, guarantee that the resulting tree  is a BT rooted at the source.  
The values of $y_{i,j}$,  found by \eqref{eq:yj}, are used in  \eqref{eq:const_d} to find the downstream values of the links between the nodes.
Eq. \eqref{eq:const_d} represents the constraint on the downstream values.
More precisely, $y_{i,j} = 0$ in \eqref{eq:const_d} indicates that, for a given $\mathbf{t}_j$,  node $i$ cannot be covered by node $j$ and the downstream value of the link between the nodes $j$ and $i$ must be zero, that is, $d_{i,j} = 0$.

Finally, it should be mentioned that the proposed MILP can also be used for the MFPBT and the MTBT problems \cite{Chekuri:JSAC07, Chen_TMC13}, however, due to the fixed transmit power, the number of constraints for the MFPBT problem will be much lower than that of the MPBT problem.
In fact, since every node has only two choices, that is, whether to transmit or not, there will be only a reachability vector for the nodes and no reachability matrix.

By the solution of the MILP, a node $j \in \mathcal{Q}$ is a transmitting node if $ \sum_{i =1}^{N} t_{i,j} =1$ and its transmit power is equal to $p_{j}^\mathrm{Tx} = p_{i,j}^\mathrm{uni}$ if $t_{i,j}=1 $.
As a node can be covered by multiple transmitting nodes in the network, an algorithm is required to find the set  $\mathcal{M}_j$ of each PN $j \in \mathcal{Q}$  in the optimum BT as well as the route $\mathcal{R}_i$ of every receiving node in $\mathcal{W}$.
To this end, we suggest Algorithm \ref{alg1}.  
In this algorithm, using the solution of the MILP and starting from the source,  node $i \in  \mathcal{W}$ is a CN of node $j$ if $y_{i,j} \neq 0$ and node $i$ has not been already connected to the BT.
The set $\mathcal{C}$ in this algorithm refers to the set of nodes which are connected to the BT.
This set at first contains $\mathrm{S}$ and the algorithm is run  until all the nodes of the network are added to this set.
The algorithm visits the nodes one by one to see if based on the solution of the MILP, a given node must be a PN of other nodes or not. 
In this algorithm, the set of visited nodes by the algorithm is given by $\mathcal{V}$.
 \begin{algorithm} [t]
 \caption{Constructing the optimum BT}
 \label{alg1}
 \begin{algorithmic}[1]
 \small
 \State $\mathcal{C} = \{\mathrm{S}\}$, $\mathcal{V} = \varnothing$   
 \While { $\mathcal{Q} \setminus  \mathcal{C} \neq \varnothing $} 
 \For{ each node $j \in \mathcal{C}\setminus \mathcal{V}$ }
 \State $\mathcal{V} = \mathcal{V} \cup j$
 \If {$y_{i,j} \neq 0, i\in \mathcal{N}_j, i \notin \mathcal{C}$ }
 \State $\mathcal{M}_j = \mathcal{M}_j \cup \{i\}$
 \State $\mathcal{R}_i = \mathcal{R}_j \cup \{i\}$
 \State  $\mathcal{C} = \mathcal{C}  \cup \{i\}$
 \EndIf
 \EndFor
 \EndWhile
  \end{algorithmic}
 \end{algorithm}
The number of binary variables that have to be determined with the proposed MILP are $N$ binary variables for the source and, as $t_{j,j} = 0$ and $t_{\mathrm{S}, j} = 0$, a number $N-1$ of binary variables for each of the $N$ nodes in $ \mathcal{W}$.
Thus, the total number of binary variables for the proposed MILP is $N + N(N-1) = N^2$.

\section{ Simulation Results} \label{sec:simul} 
 \subsection{Simulation Setup}
For simulation, a 250m$\times$250m area is considered in which the coordinate of a node is determined by $(x,y)$ with $x$ and $y$ as independently and uniformly distributed random variables in the interval [0, 250]. 
The total number of nodes varies between 10 and 50.\\
The simulation results are  based on the Monte-Carlo method and in each simulation run, one of the nodes in the network is randomly chosen as the source.
The channel is based on the path-loss model.  
Let  $l_{i,j}$ and $l_0$ be the distance between nodes $i$ and $j$ and a reference distance, respectively. Then, by considering  $\alpha$ as the path loss exponent and $\lambda$ as the signal wavelength, the power gain of the channel between nodes $i$ and $j$ is defined as 
\begin{equation}
 \label{eq:ChModel}
 g_{i,j} = \left(\frac{\lambda}{4 \pi l_0 } \right)^2 \left(\frac{l_0}{l_{i,j}}\right)^\alpha 
\end{equation}
During the simulation, we set $\lambda=0.125$m, $l_0 = 1$m and  $\alpha=3$.  
Moreover, using \cite{Cui_TWC05},  we assume uniformly distributed random values for  $p_i^\mathrm{max} \in [ 150,  250]$ mW  and  $p_j^\mathrm{c}  \in [ 50, 100]$ mW and $\eta_j = 0.3$ for all $j \in \mathcal{Q}$. 
The minimum SNR for successful decoding is set to $\gamma^\mathrm{th} = 10$ dB and the noise power is assumed to be $\sigma^2 = -90$ dBm.
We compare our algorithm with the conventional centralized and decentralized algorithms.
The benchmarks of our algorithm are the optimum solution of the MILP, explained in Sec. \ref{sec:MILP}, and  the BIPSW  \cite{BIP}, the BDP \cite{Rahnavard08} and the GBBTC \cite{Chen_TMC13}  algorithms  which are discussed in Sec. \ref{sec:intro} and Sec. \ref{subsec:discuss}.
It should be noted that the result obtained  for the MFPBT problem using the ES scheme  will be similar to that of GBBTC for the MTBT problem, on average.
Hence, we just use the GBBTC algorithm representing both.
The results for the network power for all the algorithms are normalized to the average of the maximum power budget of the nodes denoted by $\bar{P}^\mathrm{max}$, i.e., $\bar{P}^\mathrm{max} = \mathbb{E}_{j\in \mathcal{Q}}[ p_j^\mathrm{c} + p_j^\mathrm{max}$].
The normalized network power is then denoted by $\hat{P}^\mathrm{net}(\mathbf{a}) = P^\mathrm{net}(\mathbf{a})/ \bar{P}^\mathrm{max}  $ in which $P^\mathrm{net}(\mathbf{a})$ is defined in \eqref{eq:ptot}.
The simulation has been carried out in MATLAB  and the proposed MILP is solved using CVX  and Gurobi.

It should be noted that we apply no changes to the benchmark algorithms. 
For instance, in terms of the circuitry power, the benchmarks ignore it and we also implement them in this way.
After constructing the BT by those algorithms, we consider the circuitry powers in calculating the actual network power.
Modifying those algorithms in a proper way to consider the circuitry power is out of the scope of our work. 
Furthermore, we aim to emphasize  on the impact of the circuitry power which has been largely ignored by the existing algorithms and to show that the BT resulting from those algorithms are not efficient.

\subsection{Results}

 \begin{figure*} [!t]
	\begin{minipage}{.47\textwidth}
		\centering
		\includegraphics[width=.9\columnwidth]{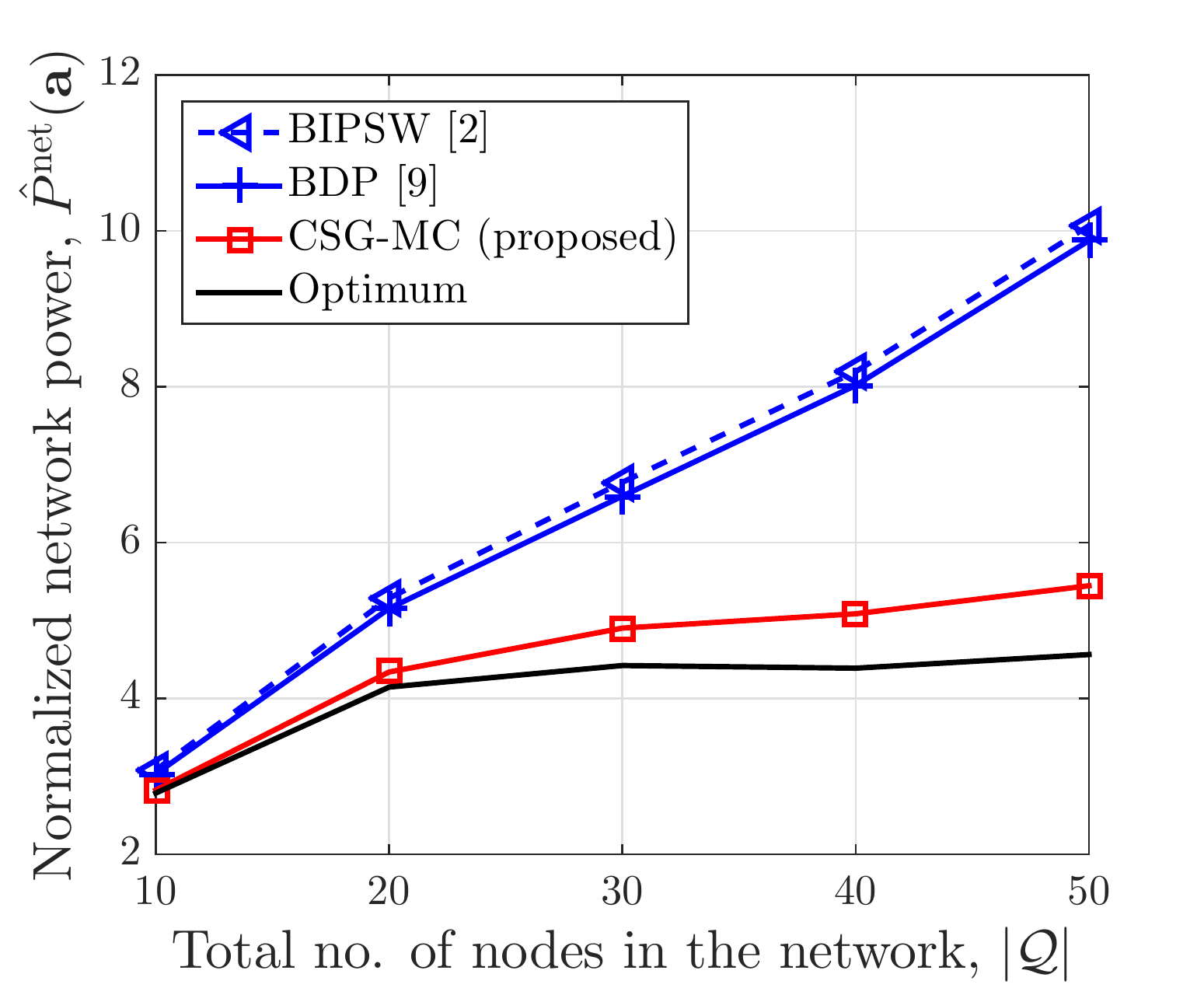}
		\caption{Normalized network power versus the number of nodes for different algorithms. }
		\label{fig:pow_all}
	\end{minipage} \qquad
	\begin{minipage}{.47\textwidth}
	\centering
  		\includegraphics[width=.9\columnwidth]{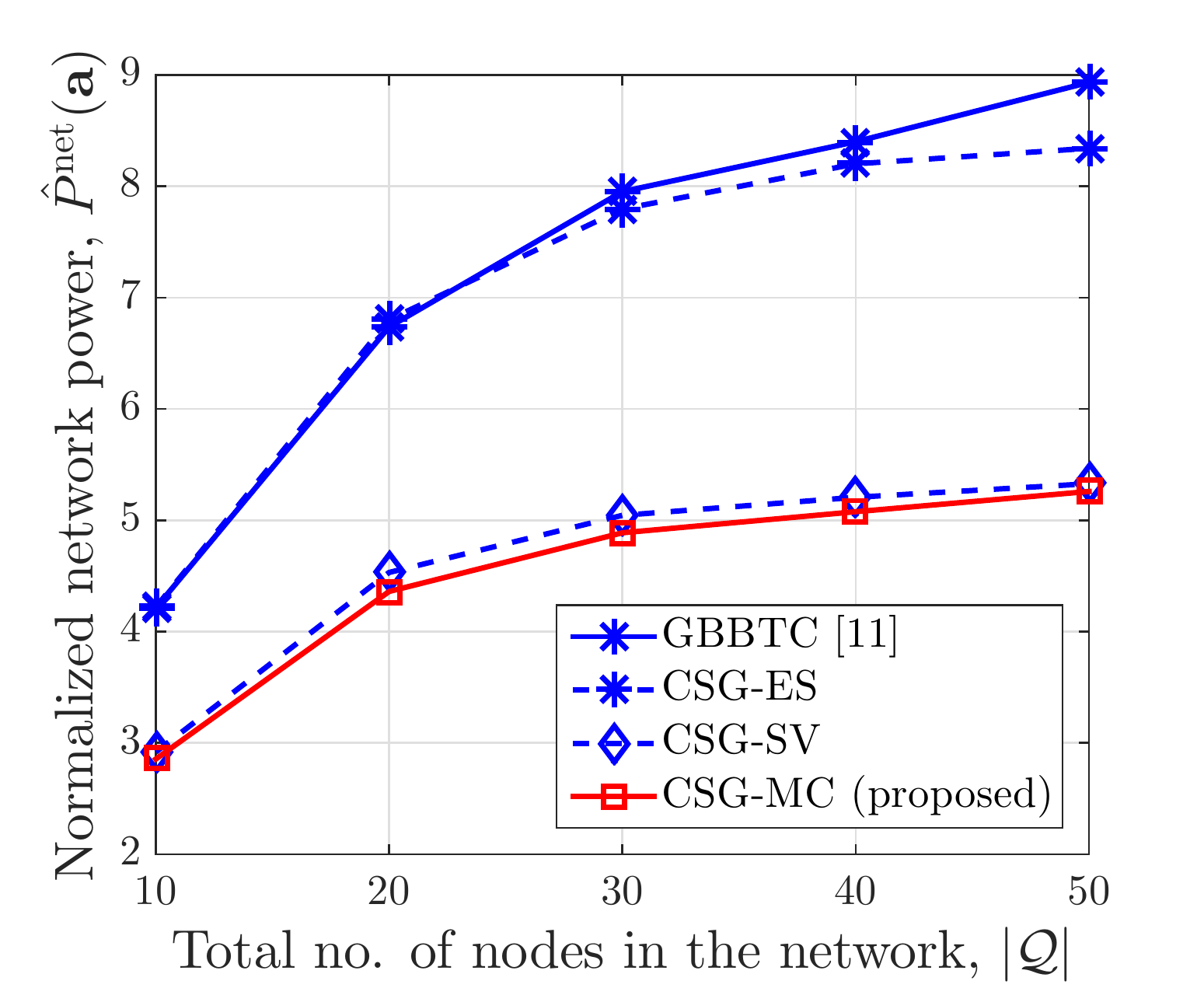}
		\caption{Comparison between different cost sharing schemes for multi-hop broadcast. }
		\label{fig:pow_MFPBT} 
	\end{minipage} 
\end{figure*}

  Fig. \ref{fig:pow_all}  compares the normalized network power  versus the number of nodes for different algorithms for the MPBT problem.
The benchmark algorithms, except the MILP, do not consider the circuitry power during the BT construction.
As can be observed, our proposed algorithm outperforms other benchmark algorithms.
The main reason is that our algorithm, besides the transmit power, considers the amount of circuitry power of the nodes and adapts the BT based on that.
In a dense network, the effect of the circuitry power on the network power is significant. 
In our algorithm,  by increasing the number of nodes, the network power first starts increasing  and then tends to saturate.
When the number of nodes in the network increases, the distances between the nodes and consequently the transmit powers required between the nodes reduce.
Despite the fact that the required transmit powers reduce,  the number of transmitting nodes in the network increases and since each transmitting node imposes a fixed power on the network, which is not negligible, the network power increases.
When the network becomes dense, the number of transmitting nodes required to cover the whole network, as well as the network power, remains roughly the same.

Fig. \ref{fig:pow_MFPBT} compares the three main cost sharing schemes discussed in this paper, that is, the MC, the SV, and the ES the in terms of the total normalized network power versus the number of nodes in the network.
We replace the MC cost sharing scheme in CSG-MC with SV and ES and refer to them as the CSG-SV and the CSG-ES, respectively.
Due to the lack of convergence guarantee with the ES scheme,  the transmit power of the nodes for the CSG-ES, as well as for the GBBTC, are assumed to be fixed and equal to 200 mW.
In this experiment, all the algorithm, except the GBBTC, consider   the circuitry power in BT construction.
In fact, the only difference between GBBTC and CSG-ES is that GBBTC relies merely on the transmit power.
There are two main observations   in Fig.  \ref{fig:pow_MFPBT}.
First, performing power control at the nodes and taking the circuitry power into account, which is the case for the CSG-MC and CSG-SV,  significantly improves the energy-efficiency of the network.
For instance, in a network with $|\mathcal{Q}|= 40$, the normalized network power obtained by CSG-MC is $\hat{P}^\mathrm{net}(\mathbf{a}) \simeq 5$.
This number for the GBBTC (an also for BIPSW in Fig. \ref{fig:pow_all}) is more than 8 which means that the BT obtained by our algorithm requires around 40\% less energy.
The second observation is that the MC   performs slightly better than  the SV.
This observation is in accordance with  Theorems \ref{th:optNE} and   \ref{th:BBNOTaligned}.
Aside from the performance, the information overhead required for the MC is much lower than that of the SV and this makes the MC the best choice for such a network.
Although the transmit power of the nodes is fixed for both the GBBTC and the CSG-ES, the network power with CSG-ES is less than that of the GBBTC.
This is because, unlike the GBBTC, the CSG-ES considers the circuitry  power in the BT formation and thus, less number of nodes act as PN.
 
\begin{figure*} [!t]
	\begin{minipage}{.47\textwidth}
		\centering
		\includegraphics[width=.9\columnwidth]{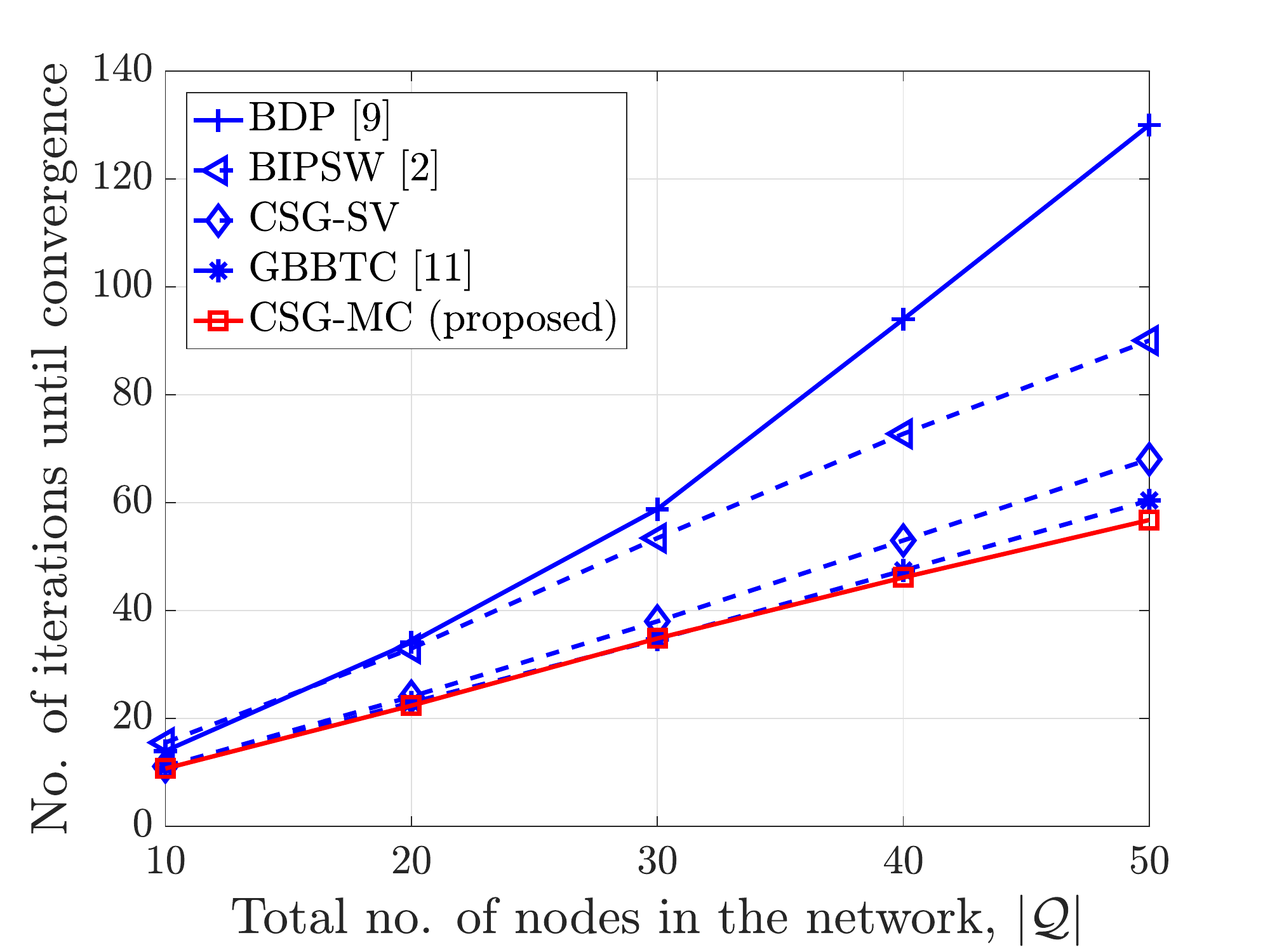}%
 		\caption{Number of iterations required for algorithms to converge. }
		\label{fig:nIter}
	\end{minipage} \qquad
	\begin{minipage}{.47\textwidth}
	\centering
  		\includegraphics[width=.9\columnwidth]{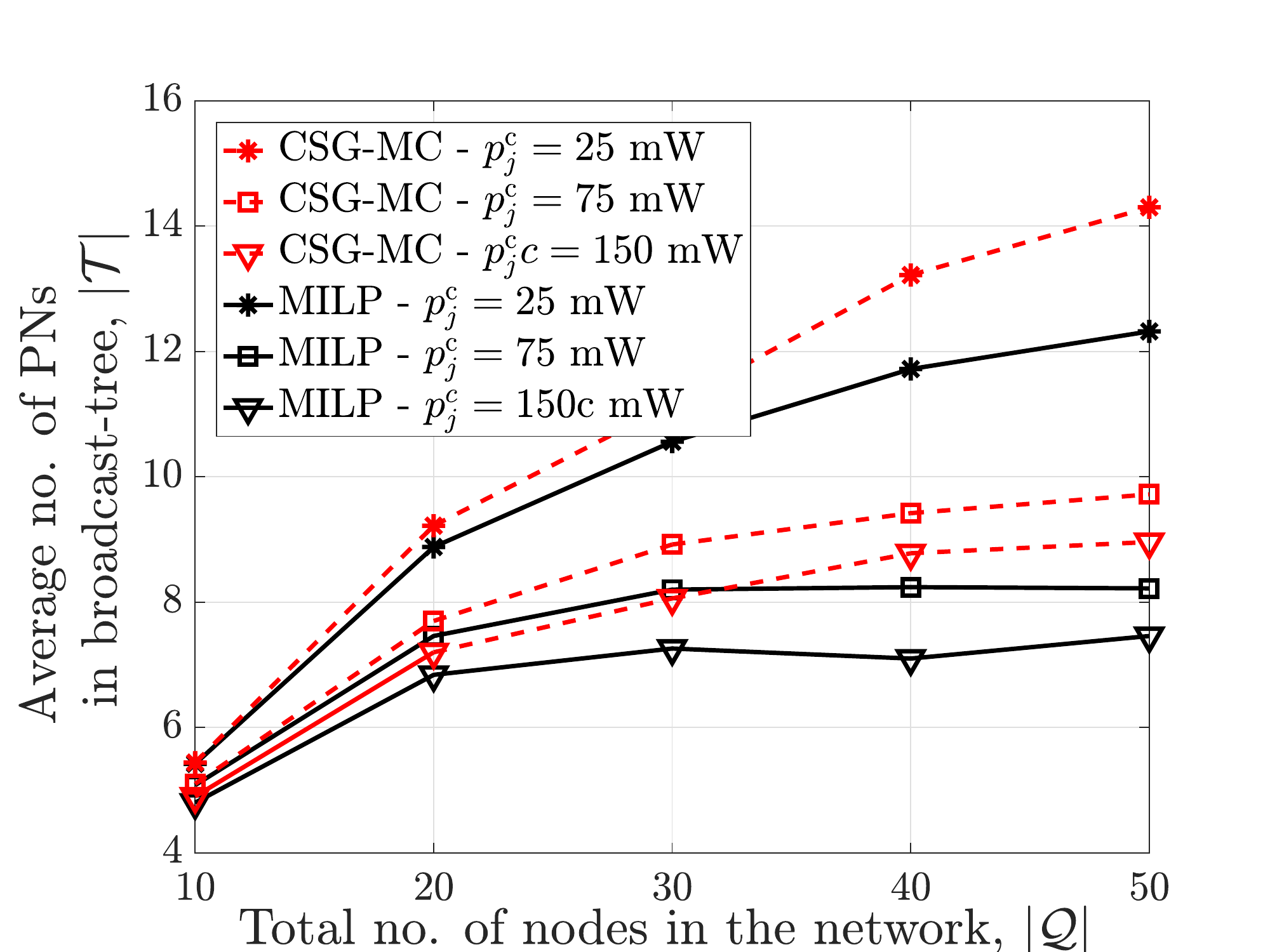}%
  		\caption{Average number of PNs (transmitting nodes) in the network for different total number of nodes. }
		\label{fig:nTx}
	\end{minipage} 
\end{figure*}

 In Fig. \ref{fig:nIter}, we depict the number of iterations required for each of these algorithms to converge.
The number of iterations  of an algorithm   can also represent its time complexity.
As can be observed, the CSG-MC algorithm requires the lowest number of iterations among all.
Moreover, the SV-based CSG requires a higher number of iterations than the MC-based CSG.
This difference stems from the way these algorithms share a cost among the receiving nodes of a multicast group.
With MC, the cost of all CNs except one of them is zero, and hence,  the CNs have no incentive to change their PN.
In contrast, the cost of the CNs with SV is  always a positive value  and the CNs may have an incentive for updating their action and finding a  PN with lower cost.
Moreover, the number of iterations required for all algorithms, except for the BDP, increases almost linearly.
The  non-linear time complexity of BDP stems from the Bellman-Ford algorithm with which the BDP needs to be initialized.
  
To show how the circuitry power affects the structure of the BT, Fig. \ref{fig:nTx} shows the average number of PNs in the network 
versus the total number $|\mathcal{Q}|$ of nodes  and for different values of the circuitry power.
It actually shows the average number of transmissions that will be carried out in the network.
The set $\mathcal{T}$ of the PNs in the network is defined as $\mathcal{T}= \{j| p_j^\mathrm{Tx} >0, j \in \mathcal{Q}\}$ where $|\mathcal{T}|$ represents the number of PNs.
As can be seen, when the circuitry power increases, the number of PNs   in the network decreases.
In this case, our proposed algorithm, as well as the  MILP,  exploit multicast transmission to reduce the network power by reducing the number of transmissions.
For instance, when  $|\mathcal{Q}|=30$ and  the average value of the circuitry power is $\mathbb{E}_{j \in \mathcal{Q}} [p_j^\mathrm{c}]=25$ mW, the BT, constructed by our algorithm, consists of roughly $|\mathcal{T}| = 11$ PNs, which means, every PN has 2.75 CNs  on average.
With $\mathbb{E}_{j \in \mathcal{Q}} [p_j^\mathrm{c}]=150$ mW, the number of PNs  becomes $|\mathcal{T}| = 8$, that is, 3.75 CNs  on average for every PN.

\begin{figure}[!t]
\begin{center}
\begin{multicols}{3}
	{ \includegraphics[width=.9\linewidth]{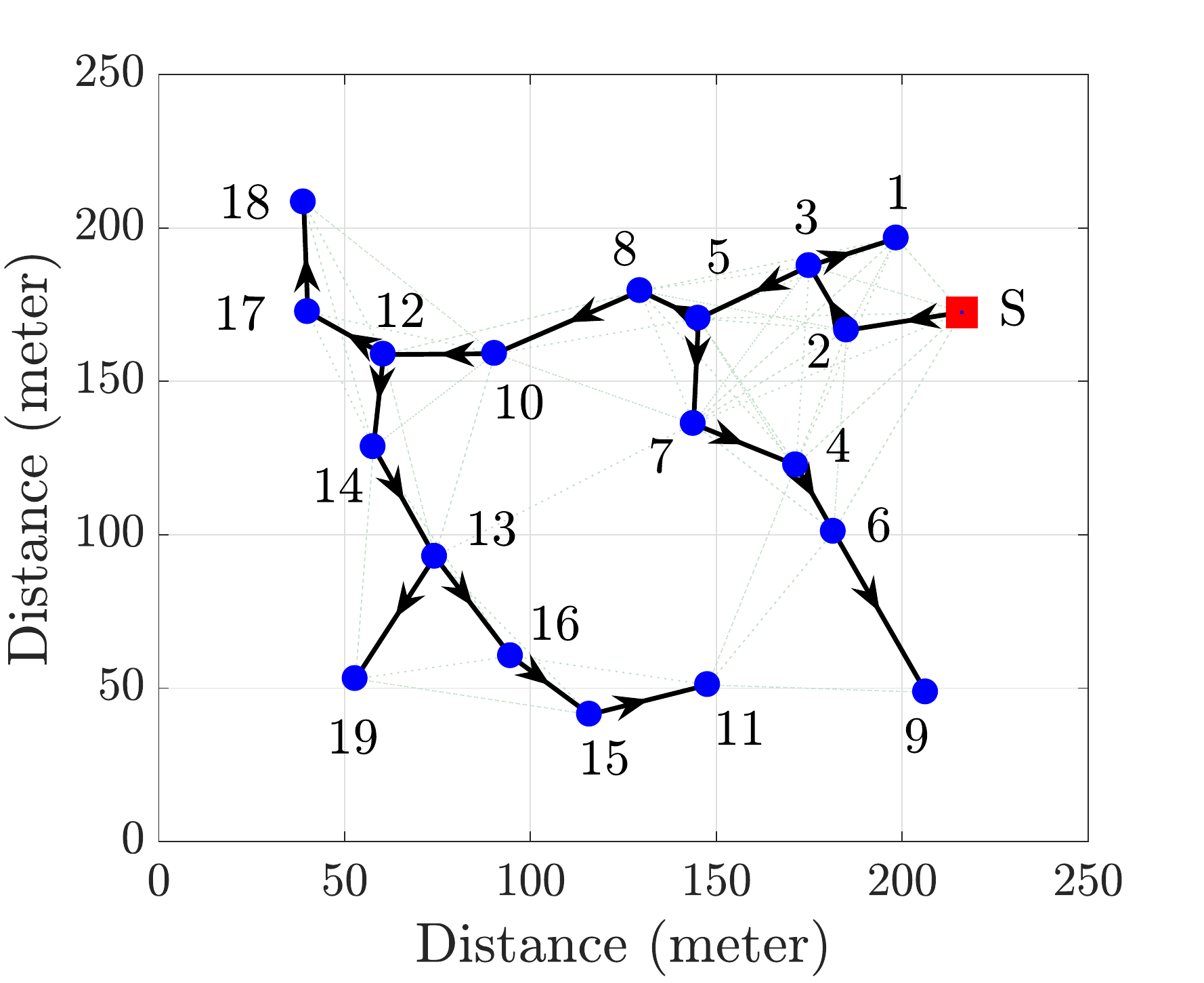} \label{fig:bt-topol-BIPSW} {\footnotesize (a) BIPSW \cite{BIP} } } \quad 	
    {\includegraphics[width=.9\linewidth]{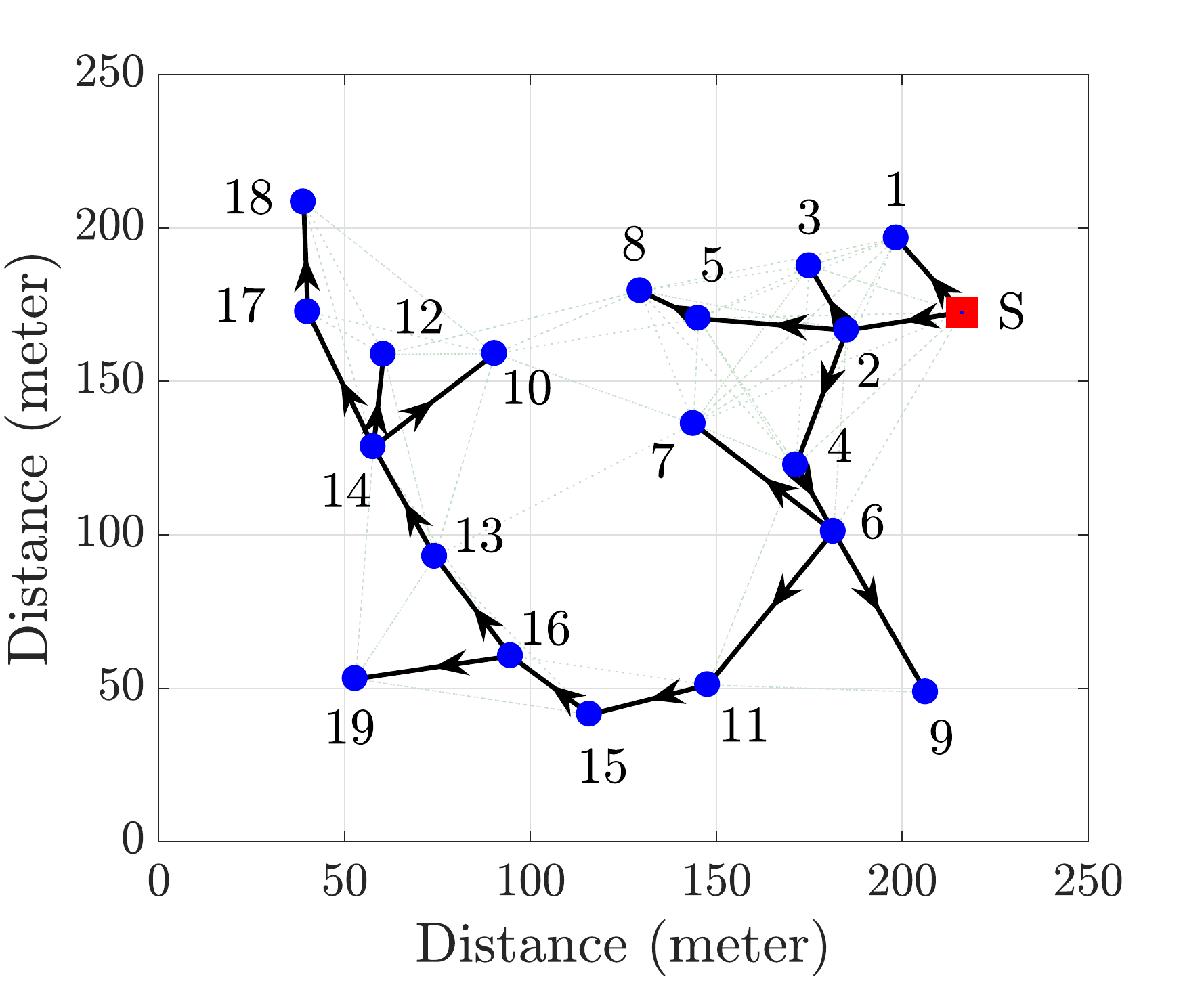} \label{fig:bt-topol-MILP25} {\footnotesize (b) Optimum - $p_j^\mathrm{c} = 25$ mW }}  \quad
    {\includegraphics[width=.9\linewidth]{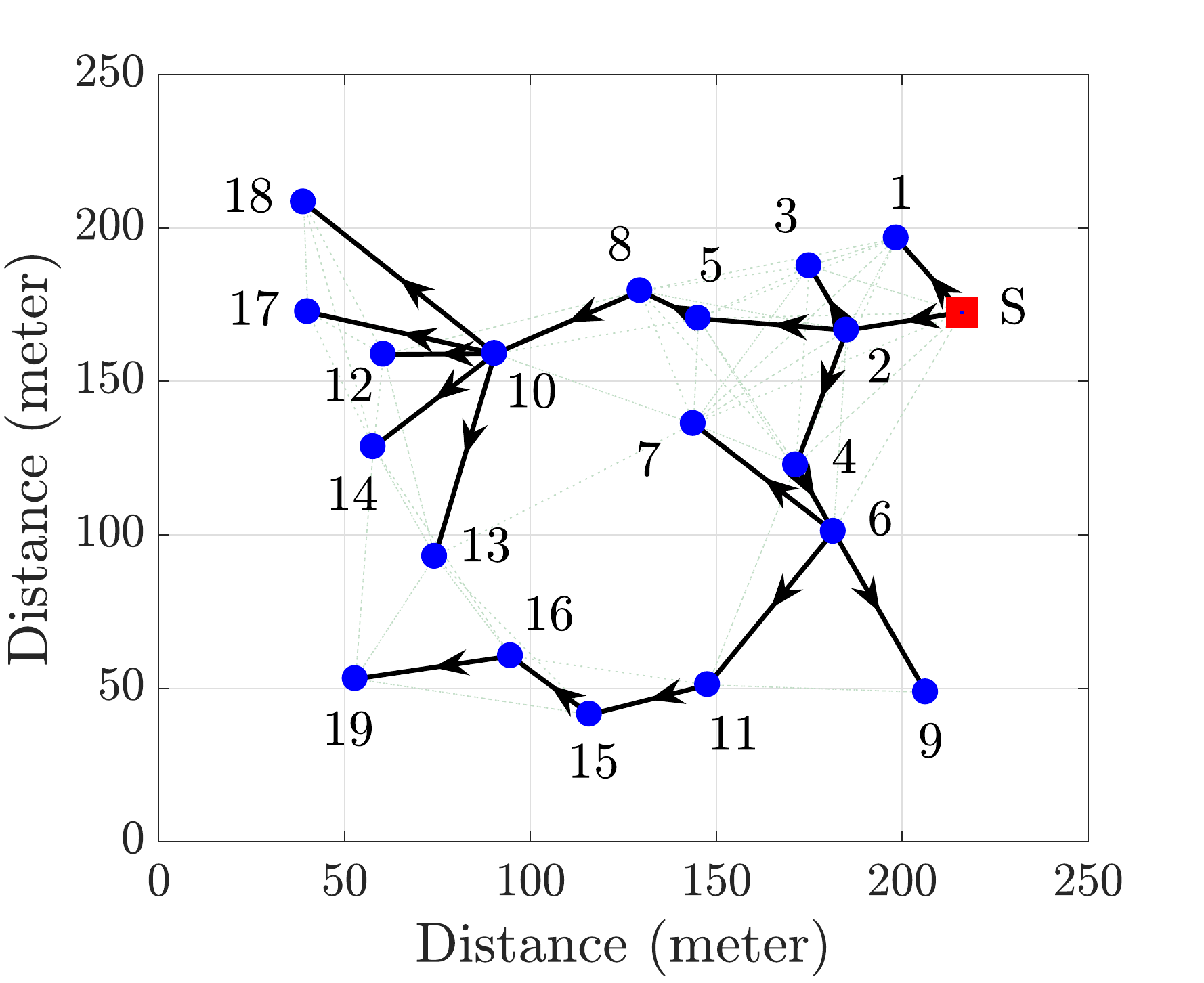}  \label{fig:bt-topol-MC25} {\footnotesize (c) Proposed  - $p_j^\mathrm{c} = 25$ mW }}\quad
\end{multicols} 
\end{center}
 \begin{center}
\begin{multicols}{3}
	{\includegraphics[width=.9\linewidth]{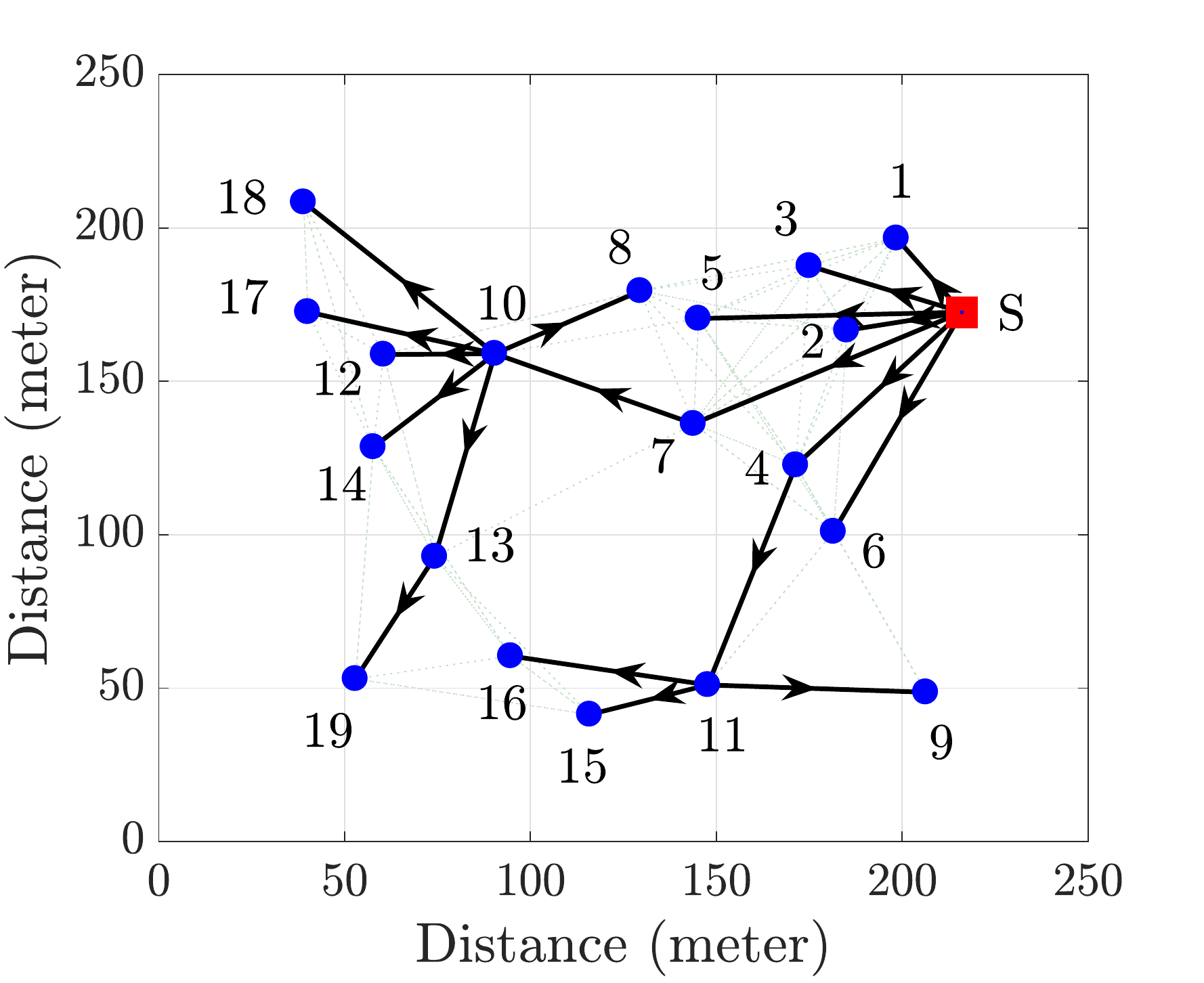} \label{fig:bt-topol-GBBTC} {\footnotesize (d) GBBTC \cite{Chen_TMC13} }} 
\quad	  
	{\includegraphics[width=.9\linewidth]{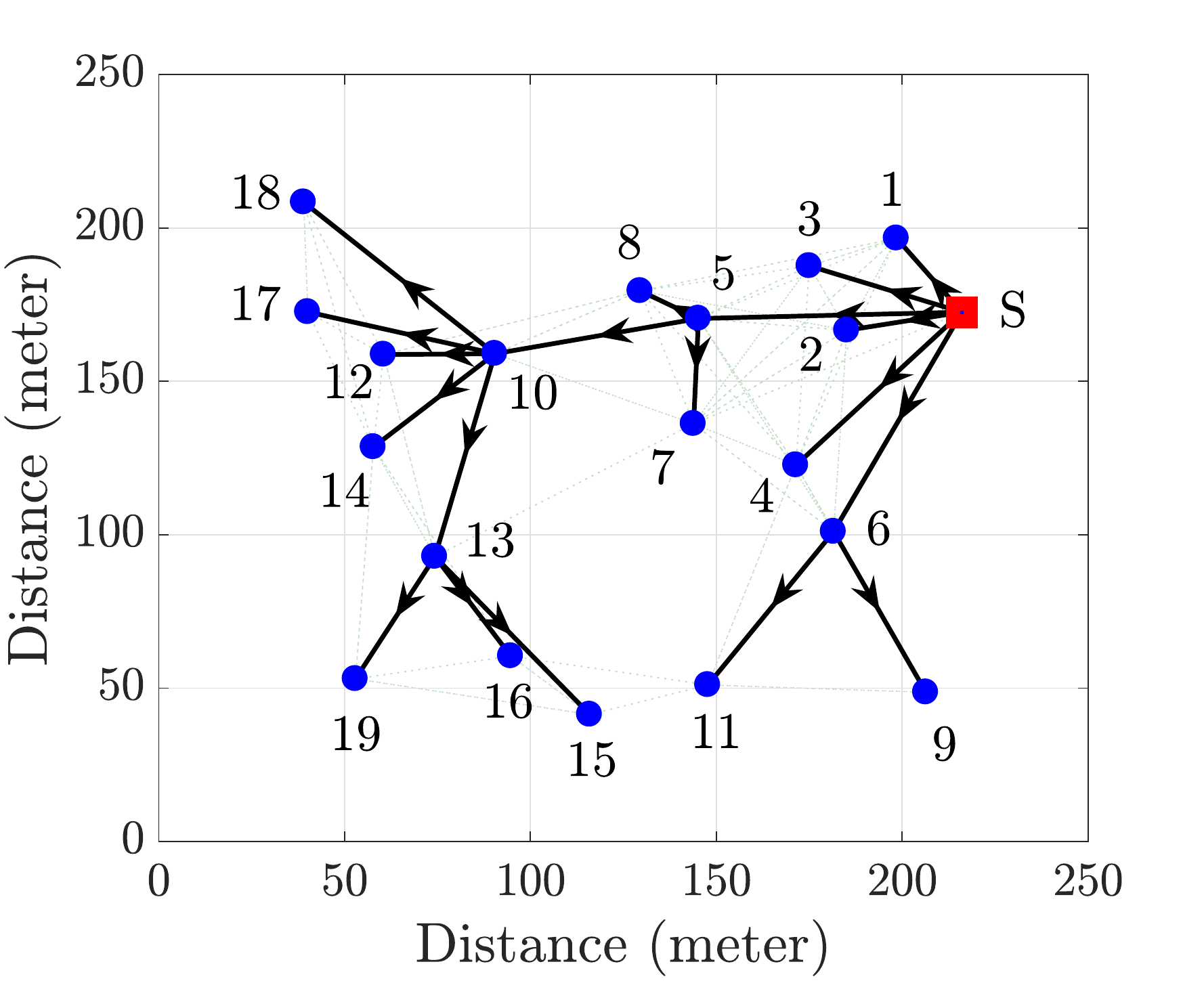} \label{fig:bt-topol-MILP150} {\footnotesize (e) Optimum - $p_j^\mathrm{c} = 150$ mW }}  
\quad	 
	{\includegraphics[width=.9\linewidth]{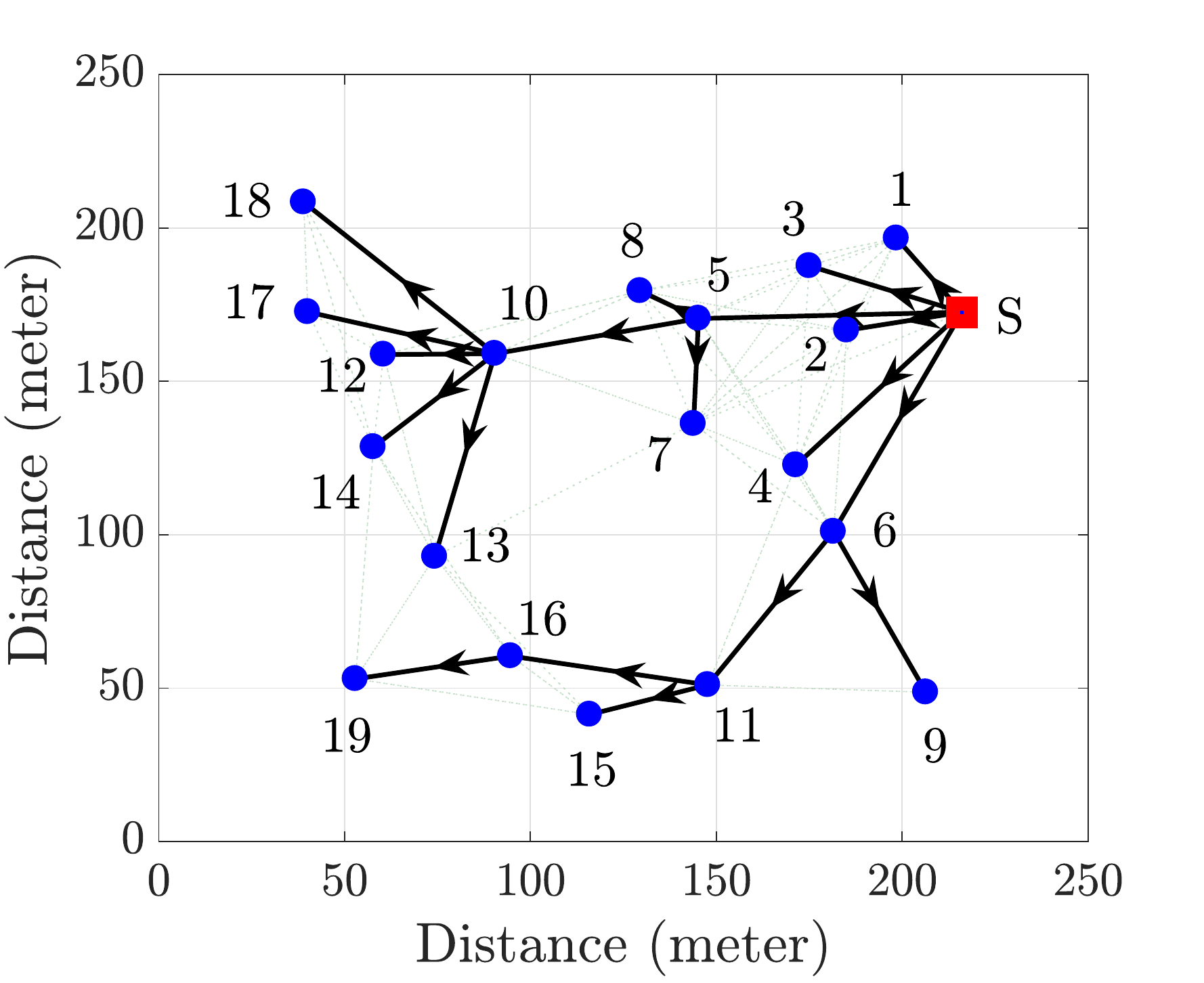} \label{fig:bt-topol-MC150} {\footnotesize (f) Proposed - $p_j^\mathrm{c} = 150$ mW }}     
\end{multicols} 
\caption{BT resulting from different algorithms in a 250m$\times$250m area with $|\mathcal{Q}| = 20$.}  
\label{fig:bt-topol}
\end{center}
\end{figure}
Finally, to have a better insight about how the algorithms construct the BT, a realization of the network with $ |\mathcal{Q}|  =20$ nodes is presented in Fig. \ref{fig:bt-topol}.
In this figure, the BT is constructed with four algorithms; the optimum BT in Fig. \ref{fig:bt-topol}~(b)  and \ref{fig:bt-topol}~(e) based on the centralized MILP approach along with the Algorithm \ref{alg1} explained in Section  \ref{sec:MILP}, the proposed decentralized game theoretic algorithm in Fig. \ref{fig:bt-topol}~(c)  and \ref{fig:bt-topol}~(f), the GBBTC \cite{Chen_TMC13} in Fig. \ref{fig:bt-topol}~(d) and  the centralized BIPSW \cite{BIP} in Fig. \ref{fig:bt-topol}~(a).
In this experiment, to show the impact of the circuitry power on the BT construction, the MILP and CSG-MC algorithms are run for two different values of the average circuitry power, that is, $  p_j^\mathrm{c} =25$ mW in (b) and (c), and $ p_j^\mathrm{c} = 150$ mW in (e) and (f) which is assumed to be the same for all the nodes $j \in \mathcal{Q}$.
Recall that the GBBTC and BIPSW ignore the circuitry power.
In Fig. \ref{fig:bt-topol} the nodes with just one outgoing link  represent the PNs that transmit via unicast while multiple outgoing links show a multicast transmission.
For instance, in the obtained BT in Fig. \ref{fig:bt-topol}~(a), node 2 receives the message from the source by a unicast transmission and sends it to its CN, i.e., node 3, again by a  unicast.
Node 3 then forwards the message to its CNs, node 1 and 5, via multicast.
In Fig. \ref{fig:bt-topol}~(a), we first find that the BIPSW constructs the BT mostly  with short hops including many unicasts.
This is because the BIPSW   relies merely on minimization of the transmit powers.
For the given instance, BIPSW requires 14 transmissions in total where 11 of these transmissions are via unicast.
In contrast to BIPSW, the GBBTC in Fig. \ref{fig:bt-topol}~(d), due to the fixed transmit power of  the nodes, tends to form large multicast groups to reduces the number of transmissions.  
 
Our proposed algorithm, as well as the optimum MILP-based BT, are  flexible.
When the circuitry power is very low,  the obtained BTs, similar to that   obtained by the BIPSW, will be  constructed  by short hops and the unicast transmission is used relatively more often.
For instance, with $p_j^c = 25$ mW, the BT constructed by our algorithm  in Fig. \ref{fig:bt-topol}~(c) contains 10 transmissions including 6 unicasts.
With the MILP in Fig. \ref{fig:bt-topol}~(b), 11 transmissions are needed with also 6 unicasts.
When the circuitry power increases to $p_j^c = 150$ mW, in the same network, the number of transmissions  with our algorithm becomes  6 including 1 unicast (Fig. \ref{fig:bt-topol}~(f)), while, the optimum BT (Fig. \ref{fig:bt-topol}~(e)) consists of 5 transmissions, all via multicast.
In fact,  when the circuitry power, as a fixed term that affects the total transmit power of a node, dominates the transmit power, our proposed algorithm as well as the MILP,  tend  to exploit the multicast transmission.
In other words, it adapts itself depending on the value of the circuitry power.

\section{Conclusion} \label{sec:conc}
In this paper, a non-cooperative cost sharing game with MC cost sharing scheme has been proposed for the MPBT problem in multi-hop wireless networks.
The proposed game has been shown to be a potential game with guaranteed convergence.
We showed that the MC cost sharing scheme is the scheme for which the optimum BT is always an NE of the game.
Besides, the information overhead required for it is relatively low.
These two properties  make it the best choice among the  cost sharing schemes for such a problem in terms of both performance and required information overhead. 
Unlike many of the existing algorithms, our proposed model not only captures the circuitry power of a device together with its transmit power but also the nodes in our algorithm are able to perform transmit power control. 
It has been shown that the proposed algorithm and the considered power model significantly improve  the network energy-efficiency.
  
\section{Acknowledgment}

This work has been funded by the German Research Foundation (DFG) as part of project B3 within the Collaborative Research Center (CRC) 1053 – MAKI. 
The authors would like to thank M. Sc. Robin Klose, working in subproject C1 of MAKI, Prof. Dr. Martin Hoefer and Prof. Dr. Yan Disser for helpful discussions.
 
\appendices
\section{} \label{app:poa}
 
\begin{figure}[!t]
\begin{center}
\begin{multicols}{2}
	{\includegraphics[width=.45 \columnwidth]{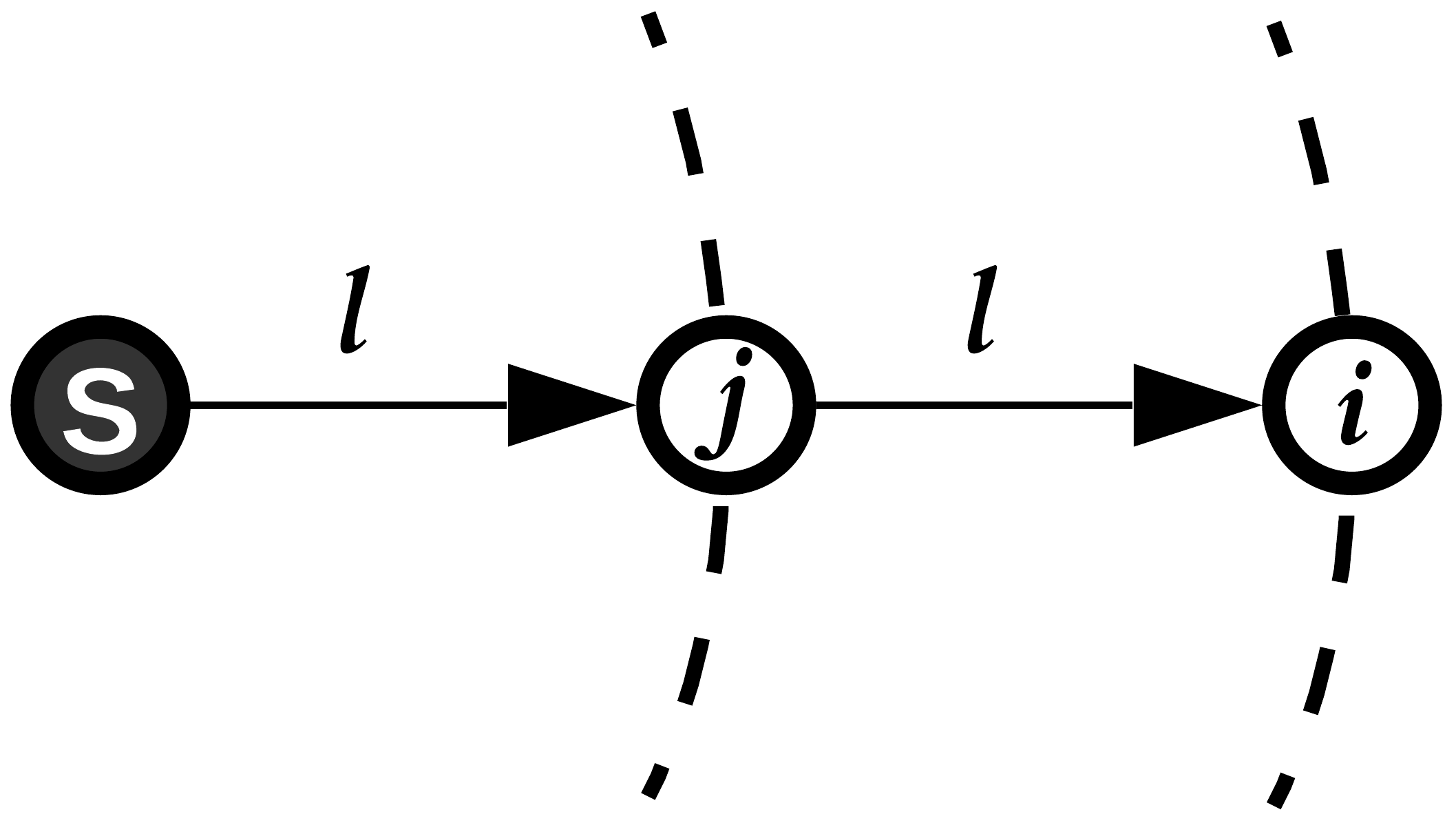} \label{fig:uni} \\ {\footnotesize (a) Unicast transmissions with $P^\mathrm{net}(\mathbf{a}_1)$} }
	\\
	{\includegraphics[width=.45 \columnwidth]{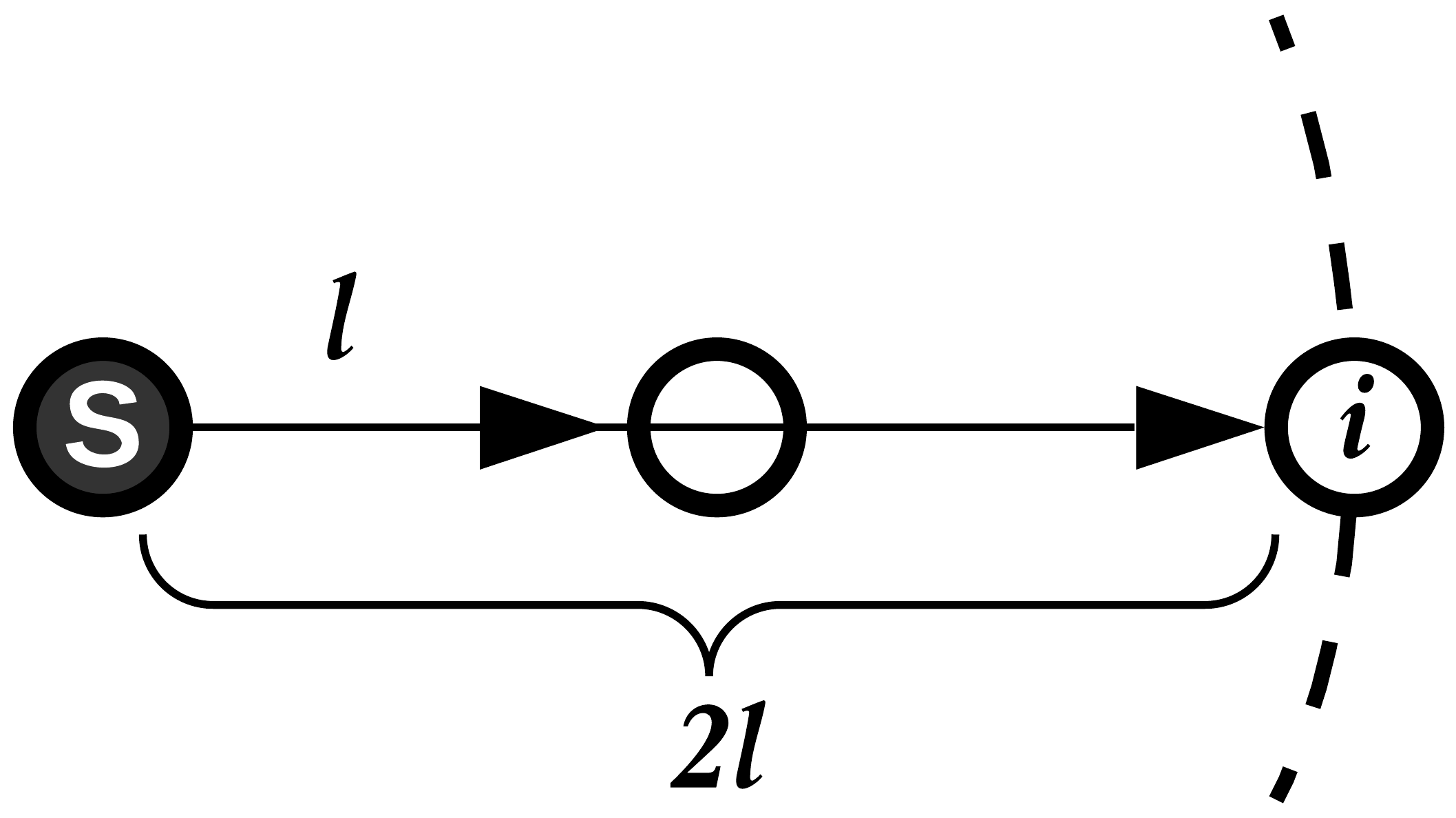} \label{fig:multi} \\ {\footnotesize (b) Broadcast with $P^\mathrm{net}(\mathbf{a}_2)$} } 
\end{multicols} 
\end{center}
\caption{Different schemes for transmission.}
\label{fig:unimulti}
\end{figure}

\begin{figure}[!t]
\begin{center}
\begin{multicols}{2}
	{\includegraphics[width=.8 \columnwidth]{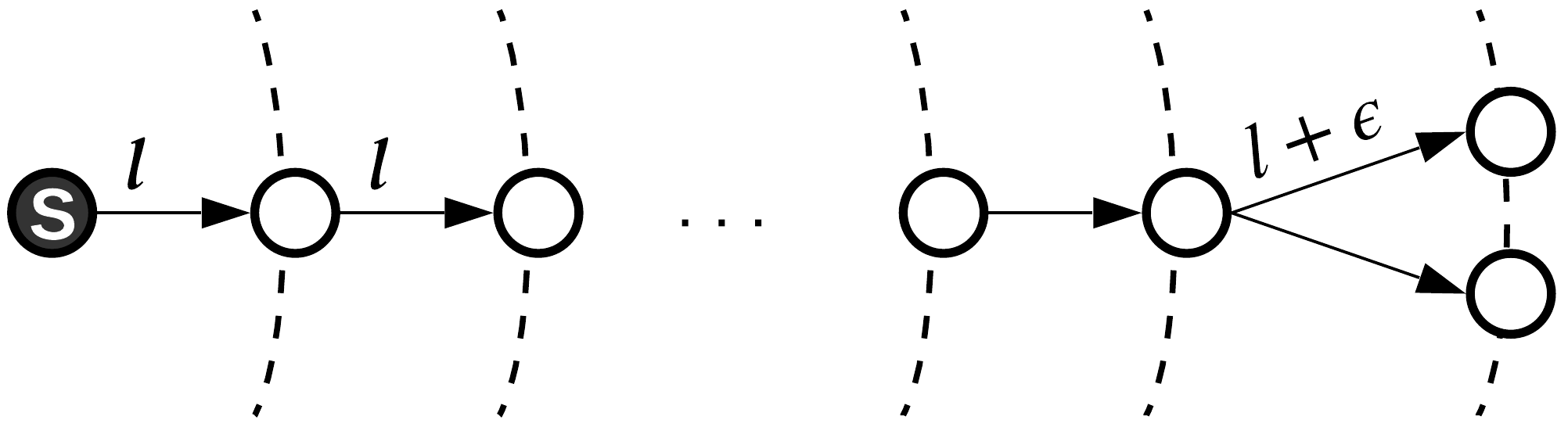} \label{fig:poa_opt} \\ {\footnotesize (a) Optimum BT with $\mathbf{a}_1$} }
	\\
	{\includegraphics[width=.8 \columnwidth]{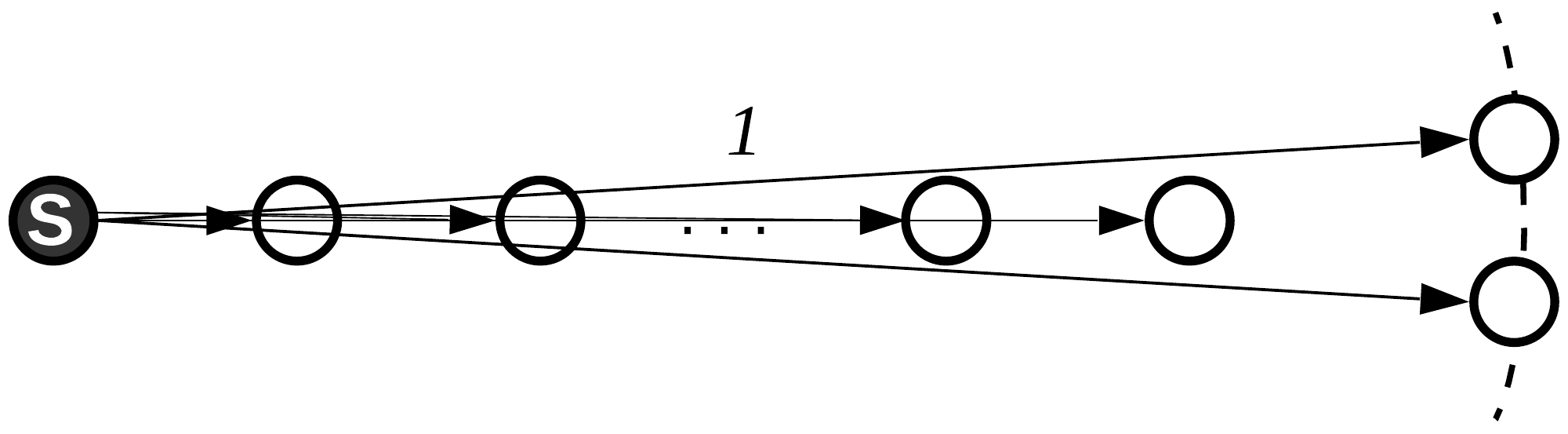} \label{fig:poa_ne} \\ {\footnotesize (b) A bad instance of NE with $\mathbf{a}_2$} } 
\end{multicols} 
\end{center}
\caption{The Optimum BT and a bad instance of NE.}
\label{fig:poa_ne_opt}
\end{figure}

In this section, we provide a lower bound for the PoA of our game,  mentioned in Theorem \ref{th:poa}.
We find an instance of an NE for which the network power compared to the global optimum is bad.
We assume a path-loss model for the channel with the path-loss exponent $\alpha$, $2 < \alpha <6 $.
Thus, the channel gain between the nodes $i$ and $j$ can be represented as $g_{i,j} \propto 1/(\tilde{l}_{i,j})^\alpha$ in which $\tilde{l}_{i,j}$ is the distance between them.
Using  \eqref{eq:puni},  the maximum transmit power is given by 
\begin{equation}
\label{p:max_d}
p_j^\mathrm{max} = \frac{\gamma^\mathrm{th} \sigma^2 \left( l^\mathrm{max} \right)^\alpha} {\eta_j} 
\end{equation}
in which $l^\mathrm{max}$ is the  largest possible distance between the PN $j$ and its CN. 
For the sake of convenience, in this section, we normalize the unicast power of the links to $p_j^\mathrm{max}$ in \eqref{p:max_d} so that the transmit power between nodes $i$ and $j$ can be represented as $p_{i,j}^\mathrm{uni} = \left(l_{i,j}\right)^\alpha$  with $l_{i,j} =\tilde{l}_{i,j}/l^\mathrm{max} $ and  $0\leq l_{i,j}\leq 1$.
Moreover,   the maximum transmit power is given by  $p^\mathrm{max} = 1$.
Further, we assume that $p_j^\mathrm{ct} =p^\mathrm{c}, \forall j \in \mathcal{Q} $.  
We now find a topology for which the game-theoretic algorithm may converge to a bad NE.
Let the nodes of the network be evenly distributed on a line as shown in Fig.~\ref{fig:unimulti} and Fig. \ref{fig:poa_ne_opt} and $l = 1/N$.
We first express the following lemma.

\begin{lemma} \label{lem:app:pc_2n}
In Fig. \ref{fig:unimulti}, $P^\mathrm{net}(\mathbf{a}_1) \leq P^\mathrm{net}(\mathbf{a}_2)$ if $p^c \leq 1-1/2^{(\alpha-1)}$.
\end{lemma}
\begin{IEEEproof}
$P^\mathrm{net}(\mathbf{a}_1) \leq P^\mathrm{net}(\mathbf{a}_2)$ if $2(p^c+l^\alpha) \leq p^c + (2l)^\alpha$ with $l = 1/2$. 
\end{IEEEproof}

\begin{lemma} \label{lem:app:pc}
Let $\mathbf{a}$ be the action profile corresponding to a BT formed merely by unicast transmissions and let $\mathbf{a}' \in \mathcal{A}\backslash \{\mathbf{a}\}$.
In the line topology shown in Fig. \ref{fig:unimulti}, similar to Lemma \ref{lem:app:pc_2n}, one can show by induction that for every $N\in \mathbb{N}$ we have $P^\mathrm{net}(\mathbf{a}) \leq P^\mathrm{net}(\mathbf{a}')$ if $p^\mathrm{c} \leq  \left( 1- 1/N^{\alpha-1} \right) / (N-2)$.
\end{lemma}
 
Let us assume that the condition in Lemma \ref{lem:app:pc} holds.
Then, given $\epsilon \rightarrow 0 $, BT in Fig.~\ref{fig:poa_ne_opt}~(a) is the optimum configuration while 
the BT in Fig.~\ref{fig:poa_ne_opt}~(b) is an NE.
The PoA given the action  profiles $\mathbf{a}_1$ and $\mathbf{a}_2$ is obtained by
\begin{equation}
\label{eq:poa_app_1}
\mathrm{PoA}(G) = \frac{ P^\mathrm{net}(\mathbf{a}_2)  }{ P^\mathrm{net}(\mathbf{a}_1) } = \frac{1 + p^\mathrm{c}}{ N \left( 1/N^\alpha  + p^\mathrm{c}\right)}.
\end{equation}
The PoA in \eqref{eq:poa_app_1} is maximized when when $p^\mathrm{c} \rightarrow 0$.
Thus, a lower bound of the PoA is  $\approx \Omega(N^{(\alpha -1)})$.

\bibliographystyle{IEEEtran}
\bibliography{ref}

\end{document}

%% file: theorem_bb_v1.tex
\begin{theorem}\label{th:BBNOTaligned}
For both ES and SV, there exists at least one instance in which $\mathbf{a}^\mathrm{opt}$ is not an NE. 
\end{theorem}

\begin{IEEEproof}
We provide an instance in Fig. \ref{fig:global-local} for which the optimum action profile is not an NE for the ES and the SV.
In this instance, using the ES,  node $i$ updates its action from $a_i = j$ to $a_i' = k$ to reduce its cost from $(p_j^\mathrm{c}+6)/2  $ to $(p_k^\mathrm{c} + 3)/2 $.
Assuming $p_j^\mathrm{c} = p_k^\mathrm{c}$, this action reduces the cost of node $i$ by 1.5 units while at  the same time, it deviates from the optimum action profile $(\mathbf{a})$ and increases the network power  by 1 unit, that is,  from $P^\mathrm{net}(\mathbf{a}) = P_\mathrm{S}(\{j,k\}) + p_j^\mathrm{c} + 6 + p_k^\mathrm{c} +1$ to $P^\mathrm{net}(\mathbf{a}') = P_\mathrm{S}(\{j,k\}) + p_j^\mathrm{c} + 5 + p_k^\mathrm{c} +3$.
Using the SV defined in \eqref{eq:SHAP-def} also  leads to the same conclusion as  node $i$ reduces its cost from 3.5 to 2.5 by the same action.
It should also be remarked that by employing the MC  in this example, node $i$ does not change its action  since such an action increases its cost from 1 to 2. 
\end{IEEEproof}
 
\begin{remark}\label{rem:bb_not_unq}
Although for the instance provided in Fig. \ref{fig:global-local} both the ES and the SV are not able to reach the optimum configuration, the instances for which $\bm{a}^\mathrm{opt}$ is not an NE may be different ones for the ES and the SV.
\end{remark}

We now discuss how can one verify if a budget-balanced cost sharing scheme is not able to reach the global optimum in general.

\begin{remark}\label{rem:bb_relation}
Let node $i$ change its action from  $a_i = j, a_i\in \mathbf{a}$ to $a_i' = k, a_i'\in \mathbf{a}'$ under a given budget-balanced cost sharing scheme $C^\mathrm{BB}$.
Then,  $\bm{a}^\mathrm{opt}$ cannot be guaranteed to be an NE for the  $C^\mathrm{BB}$ if one can find $\bm{a}$ and $\bm{a}^\prime$  for which the following holds:
\begin{equation}
\label{eq:bb_aligned_cond}
\underset{\mathbf{a} \rightarrow \mathbf{a}'}{\Delta^i} \quad
\smashoperator[lr]{ \sum_{m \in \mathcal{M}_k \cup  \mathcal{M}_j \backslash \{i\} }} C_m^\mathrm{BB}
>
- \underset{\mathbf{a} \rightarrow \mathbf{a}'}{\Delta^i} C_i^\mathrm{BB}.
\end{equation}
 
This can be shown as follows.
Using Definition \ref{def:BB} and by a summation over all the nodes $j\in \mathcal{Q}$, for a budget-balanced cost sharing scheme one can write
\begin{equation}
\label{eq:fair_1}
 \sum_{j \in \mathcal{Q}} \sum_{i \in \mathcal{M}_j} C_i^\mathrm{BB}(j,\mathcal{M}_j) =  \sum_{j \in \mathcal{Q}}   P_j (\mathcal{M}_j).
\end{equation}

The left side of \eqref{eq:fair_1} represents  the total payment \textit{received}  by the PNs $j \in \mathcal{Q}$ which is equal to the cost \textit{paid} by the CNs $i \in \mathcal{W}$, i.e., $\sum_{i \in \mathcal{W}}  C_i^\mathrm{BB}( \mathbf{a} )$. 
Thus, \eqref{eq:fair_1} is equivalent to
\begin{equation}
\label{eq:fair_2}
 \sum_{i \in \mathcal{W}} C_i^\mathrm{BB}(\mathbf{a} ) =  \sum_{j \in \mathcal{Q}}  P_j(\mathbf{a} ).
\end{equation}
By expanding the left side of \eqref{eq:fair_2} and re-arranging it, the cost of node $i$ is given by
\begin{equation} 
\label{eq:fair_3}
   C_i^\mathrm{BB}(\mathbf{a} )  =   \sum_{j \in \mathcal{Q}}  P_j(\mathbf{a} )  - \smashoperator[lr]{ \sum_{m \in \mathcal{W} \backslash \{i\} }} C_m^\mathrm{BB}(\mathbf{a} ) = P^\mathrm{net}(\mathbf{a}) -\smashoperator[lr]{ \sum_{m \in \mathcal{W} \backslash \{i\} }} C_m^\mathrm{BB}(\mathbf{a} )
\end{equation}
in which $P^\mathrm{net}(\mathbf{a})$ is defined in \eqref{eq:ptot}.
Since in  transition from  $a_i = j$ to $a_i' = k$,   only the PNs $j$ and $k$ and their CNs are affected, then using \eqref{eq:fair_3} we get
\begin{equation} 
\label{eq:delata_ai_exp}
\underset{\mathbf{a} \rightarrow \mathbf{a}'}{\Delta^i} C_i^\mathrm{BB} 
=
\underset{\mathbf{a} \rightarrow \mathbf{a}'}{\Delta^i}
P^\mathrm{net}
-
\underset{\mathbf{a} \rightarrow \mathbf{a}'}{\Delta^i}
\smashoperator[lr]{ \sum_{m \in \mathcal{W} \backslash \{i\} }} C_m^\mathrm{BB}  
= 
\underset{\mathbf{a} \rightarrow \mathbf{a}'}{\Delta^i}
P^\mathrm{net}
- 
\underset{\mathbf{a} \rightarrow \mathbf{a}'}{\Delta^i} \quad
\smashoperator[lr]{ \sum_{m \in \mathcal{M}_k \cup  \mathcal{M}_j \backslash \{i\} }} C_m^\mathrm{BB}
.
\end{equation}

Since $\mathbf{a} \overset{i}{\rightarrow} \mathbf{a}'$ implies that 
  $\underset{\mathbf{a} \rightarrow \mathbf{a}'}{\Delta^i} C_i^\mathrm{BB}  <0$, based on \eqref{eq:delata_ai_exp},
$\underset{\mathbf{a} \rightarrow \mathbf{a}'}{\Delta^i}
P^\mathrm{net}
 < 
\underset{\mathbf{a} \rightarrow \mathbf{a}'}{\Delta^i}
\smashoperator[lr]{ \sum_{m \in \mathcal{M}_k \cup  \mathcal{M}_j \backslash \{i\} }} C_m^\mathrm{BB} 
$
 which does not necessarily indicate $\underset{\mathbf{a} \rightarrow \mathbf{a}'}{\Delta^i} P^\mathrm{net} <0$.
More precisely, based on \eqref{eq:delata_ai_exp},  if  the condition in \eqref{eq:bb_aligned_cond} is met, then, $\underset{\mathbf{a} \rightarrow \mathbf{a}'}{\Delta^i}
P^\mathrm{net} > 0$.
Roughly speaking, according to \eqref{eq:bb_aligned_cond}, if the increase in the sum of the costs of the CNs in $m \in \mathcal{M}_k \cup  \mathcal{M}_j \backslash \{i\}$ is higher than the reduction that CN $i$ experiences in its cost, then the network power increases by $\mathbf{a} \overset{i}{\rightarrow} \mathbf{a}'$.

Hence,  if one finds an instance of the network for which \eqref{eq:bb_aligned_cond} holds, then, the cost of a CN and the network power are not aligned under $C^\mathrm{BB}$ and deviating from an action profile  $\mathbf{a} \in \mathcal{A}$ does not, in general, result in network power reduction.
Since   $ \mathbf{a}^\mathrm{opt} \in \mathcal{A}$, the global optimum cannot be guaranteed to be an NE for $C^\mathrm{BB}$.
\end{remark}